\documentclass[10pt,twocolumn,twoside]{IEEEtran}

\usepackage{graphicx,cite}
\usepackage{array,amssymb,amsmath,amsfonts}
\usepackage{subfigure}
\usepackage{multirow}
\usepackage{math}

\usepackage{widetext}

\usepackage{bm,color}

\title{Hyper-Spectral Image Analysis with Partially-Latent Regression and Spatial Markov Dependencies}
\author{
\IEEEauthorblockN{{Antoine Deleforge}\IEEEauthorrefmark{2}} \and
\IEEEauthorblockN{{Florence Forbes}\IEEEauthorrefmark{1}} \and
\IEEEauthorblockN{{Sileye Ba}\IEEEauthorrefmark{1}} \and
\IEEEauthorblockN{{Radu Horaud}\IEEEauthorrefmark{1}}\\
\IEEEauthorblockA{\IEEEauthorrefmark{1} INRIA Grenoble Rh\^one-Alpes, Montbonnot Saint-Martin, France}\\
\IEEEauthorblockA{\IEEEauthorrefmark{2} Friedrich-Alexander-Universit\"at, Erlangen-N\"urnberg,
91058 Erlangen, Germany}
\thanks{The authors acknowledge support from 
the European Research Council through the ERC Advanced Grant VHIA \#340113.}
}

\usepackage[final]{review} 

\begin{document}
\markboth{IEEE Journal of Selected Topics in Signal Processing, vol.~9, no.~9, September~2015}%
{Deleforge \MakeLowercase{\textit{et al.}}: Hyper-Spectral Image Analysis with Partially-Latent Regression and Spatial Markov Dependencies}
\maketitle
\begin{abstract}
Hyper-spectral data can be analyzed to recover physical properties at large planetary scales. This involves resolving inverse problems which can be addressed within  machine learning, with the advantage that, once a relationship between physical parameters and spectra has been established in a data-driven fashion, the learned relationship can be used to estimate physical parameters for new hyper-spectral observations. Within this framework, we propose  a spatially-constrained and partially-latent regression method which maps high-dimensional inputs (hyper-spectral images) onto low-dimensional responses (physical parameters \addnote[physical1]{1}{such as the local chemical composition of the soil}).  The proposed regression model comprises two key features. Firstly, it combines a Gaussian mixture of locally-linear mappings (GLLiM) with a partially-latent response model. While the former makes high-dimensional regression tractable, the latter enables to deal with physical parameters that cannot be observed or, more generally, with data contaminated by experimental artifacts that cannot be explained with noise models. Secondly, spatial constraints are introduced in the model through a Markov random field (MRF) prior which provides a spatial structure to the Gaussian-mixture hidden variables. Experiments conducted on a database composed of remotely sensed observations collected from the Mars planet by the Mars Express orbiter demonstrate the effectiveness of the proposed model. 

\end{abstract}
\section{Introduction}
\label{sec:introduction}
In modern geosciences, the use of remotely sensed data acquired from airborne, spacecraft, or satellite 
sensors allows to acquire multimodal observations at planetary levels \cite{Bibring2004,ChangOceanography2004,CloutisWHISPERS2009,FionaHiltonBAMS2012}. Nowadays, hyper-spectral observations
are used to monitor geophysical and environmental phenomena characterized by complex spectral signatures covering wide spectrum ranges.

Usually, recovering physical parameters \addnote[physical2]{1}{such as the local chemical composition of the soil}  from hyper-spectral images involves solving inverse problems. This typically requires building complex high-dimensional to low-dimensional mappings relating the hyper-spectral space to the physical parameter spaces. Many approaches can be considered such as physically-based models, which build high-to-low dimensional mappings from 
geophysical models  \cite{kimes2000RSR,GoodenoughIGARSS2006,Stenberg2008ALRS,ShokrIGARSS2003,FosterJoG2012}.
These models follow the laws of physical phenomena and are driven by cause-to-effect relationships.
In \cite{kimes2000RSR}, vegetation reflectance models are built to extract vegetation variables from satellite sensors directional and spectral information. 
\cite{GoodenoughIGARSS2006,Stenberg2008ALRS} integrate hyper-spectral observations to a canopy radiative transfer model for forest species classification, and forest biophysical and biochemical properties estimation. 
\cite{ShokrIGARSS2003,FosterJoG2012} use multi-spectral observations from multiple sensors, combined with physical radiative transfer models, to estimate ice surface characteristics such as temperature, concentration, and thickness.
There are issues related to physically-based models, namely: expert knowledge is required for their construction. Moreover, due to  strong non-linear interactions between a large number of variables, their inversion is not straightforward. Indeed, inverse problems are in general ill-posed in the presence of real data. Thus, every model requires a carefully designed inversion procedure.

Another family of approaches learn input-output mappings  in a data driven fashion. These approaches have several advantages. Firstly, they do not require complete knowledge of the underlying physical phenomena. Therefore, they can be applied whenever training data and reliable models are available. Secondly, learned models can be readily applied to newly sensed data \cite{bernard2009retrieval}. Furthermore, in situations where expert knowledge is available, learning based approaches can benefit from physics-based models, \textit{e.g.}, by exploiting training data produced by radiative transfer models \cite{Doute2007}. Machine learning methods are very popular for 
hyper-spectral data classification in the case of discrete-valued output variables. The reader is referred to \cite{FauvelIEEEProceedings2013} for an up to date review of state-of-the-art classification models applied to hyper-spectral data. Recently, \cite{ChenIEEEJST2014} proposed a deep learning approach for hyper-spectral data classification. When the output variables are continuous,  existing regression methods are mainly concerned with hyper-spectral data un-mixing where the goal is to identify end-members constituents  together with their corresponding abundance from hyper-spectral observed mixtures  \cite{BhattIEEEJST2014,HeylenIEEEJST2014}.

In the machine learning literature, regression models have been extensively studied. When dealing with high-dimensional input data, such as hyper-spectral observations, the vast majority of methods start by reducing the dimensionality followed by regression \cite{Li91,wu2008kernel,AdragniCook2009}. Such two-step approaches cannot be conveniently expressed in terms of a single optimization problem. Moreover, this presents the risk to map the
input onto an intermediate low-dimensional space that
does not necessarily contain the information needed to correctly predict the
output. To estimate non-linear mappings, mixtures of locally linear models were proposed \cite{XuJordanHinton95,QiaoMinematsu09,ingrassia2012local}, but not in the case of high-dimensional inputs and in the presence of partially-observed responses. An alternative popular approach consists in the use of kernel functions, with the drawbacks that these functions cannot be appropriately chosen automatically, and that the learned mappings cannot be inverted \cite{tipping2001sparse,lawrence2005,wu2008kernel,thayananthan2006multivariate}. 

In this paper we propose a learning based regression model that maps high-dimensional hyper-spectral observations onto low-dimensional geophysical parameters. The proposed model generalizes the Gaussian mixture of locally-linear mapping (GLLiM) model recently proposed \cite{DeleforgeForbesHoraud2014}. The design principles of GLLiM consist of first, learning the parameters of an \textit{inverse} low-to-high dimensional probabilistic mixture of locally-linear \textcolor{black}{affine transformations}, and second, deriving the \textit{forward} posterior conditional density of the response, given an observed input and the learned model parameters.
Exchanging the roles of the input and of the response variables during training, makes high-dimensional regression tractable and the \textcolor{black}{conditional} density of the response has a closed-form solution. Moreover, the model is designed such that it can deal with a partially-latent response: the vector-valued response variable is composed of both observed entries and latent (unobserved) entries. This enables to handle in a principled fashion the presence, in the training data, of physical parameters and experimental artifacts that cannot be explicitly measured.

With respect to \cite{DeleforgeForbesHoraud2014}, the main novelty of the model proposed in this paper is the introduction of  a Markov random field (MRF) prior on the hidden (assignment) variables of the mixture model. This prior explicitly enforces spatial regularity of the reconstructed geophysical fields. Our model integrates high-dimensional to low-dimensional regression and MRF spatial regularization into a unified Bayesian framework.
Because model parameter learning is intractable due to the complex Markov interactions between the variables, in particular because of the MRF prior, the parameters are estimated via an approximate variational Bayesian learning scheme. We provide a detailed description of a variational EM (VEM) procedure that is used for learning the model parameters from training data.
Our modeling is quite original as in general MRF priors are enforced onto response variable fields \cite{TarabalkaIEEEGSRL2010,BhattIEEEJST2014}. Furthermore, because the mixture's hidden variables are discrete and finite, while response variables are continuous, our setting is computationally more tractable. 
\addnote[addbiblio]{1}{Another interesting feature is that the Markov field is  directly included in the generative modeling of the high-to-low dimensional mapping. In general, such dimension reduction aims at representing spatial vectors in a lower more tractable space in which noise can be better characterized. Indeed, the dimension of the space spanned by spectra from an image is generally much lower than the available number of bands. When the goal is to characterize observed spectra with a limited number of features (such as abundances or physical characteristics), identifying appropriate subspaces can then significantly improve performance.   This reduction step is then often an intermediate step before additional refinement such as the inclusion of contextual information is performed. Typically, in \cite{Li2012}, dimension reduction and Markov modeling are carried out in two successive steps and assuming a linear mixing model. This is quite different from our approach which aims at modeling and identifying non-linear mappings and not necessarily in the un-mixing context only.  In our work,  the high-to-low mapping is a target  while it is only a convenient intermediate step in most studies. Of course, considering the high level of activity in hyper-spectral image analysis, it would be challenging to present an exhaustive view of the subject but we believe that methods using subspace reduction in this field correspond to a different use of dimensionality reduction. We refer to \cite{Bioucas-Dias2012} for one recent review on both subspace identification and spatial modeling.}

The model that we propose is also related to hyper-spectral data un-mixing \cite{Bioucas-Dias2012,Altmann-Dias2012,BhattIEEEJST2014} since in our model the constituting components of the hyper-spectral data are segregated with a locally-affine regression model. However and in contrast to existing methods, our model, through the presence of the partially-latent output, is able to handle the presence of non identifiable components in the input data. This particular property  could be interpreted as an ability to partially un-mix hyper-spectral observations while letting a remaining part unexplained.

The effectiveness of the proposed model is assessed using datasets of simulated and real hyper-spectral images of the OMEGA imaging spectrometer 
on board of the Mars Express spacecraft \cite{Bibring2004}. Thorough quantitative and qualitative results demonstrate that our model outperforms several state-of-the-art regression models. Furthermore, the fact that we integrate an MRF prior produces spatially-regular parameter maps.

The remainder of this paper is organized as follows. Section~\ref{sec:gllim} summarizes the high-dimensional partially-latent regression model \cite{DeleforgeForbesHoraud2014}. Section~\ref{sec:variational-EM} presents the novel MRF spatially-constrained regression model. Section~\ref{sec:applications} applies the proposed method to hyper-spectral images collected from Mars with the OMEGA instrument. 
Finally,  Section~\ref{sec:conclusions} draws conclusions and discusses possible future research challenges.


\section{Inverse Regression with Partially-Latent Response Variables}
\label{sec:gllim}
In this section, we summarize the inverse regression model proposed in \cite{DeleforgeForbesHoraud2014}. This model maps low-dimensional variables $\Xvect= \{\Xvect_n\}^{N}_{n=1}$ onto high-dimensional variables $\Yvect=\{\Yvect_n\}_{n=1}^N$. To each low dimensional variable 
$\Xvect_n \in \mathbb{R}^L$ corresponds a high dimensional variable $\Yvect_n \in  \mathbb{R}^D$ with $L\ll D$. 
We consider the \textcolor{black}{ {\it hybrid\footnote{hybrid refers to the observed and latent structure of the response variable $\Xvect_n$.} Gaussian locally-linear mapping }(hGLLiM)} framework proposed in \cite{DeleforgeForbesHoraud2014} that treats each $\Xvect_n$ as a partially latent variable, namely: 
\begin{equation}
\label{eq:partial-latent}
\Xvect_n=\left[
  \begin{array}{c}
   \Tvect_n \\
   \Wvect_n
  \end{array}
 \right],
\end{equation}
where $\Tvect_n\in\mathbb{R}^{L_{\textrm{t}}}$ is observed and
$\Wvect_n\in\mathbb{R}^{L_{\textrm{w}}}$ is latent, with
$L=L_{\textrm{t}}+L_{\textrm{w}}$. 
Let us consider $\{\xvect_n\}_{n=1}^{n=N}$ and  $\{\yvect_n\}_{n=1}^{n=N}$, 
where $(\yvect_n,\xvect_n)$ is a realization of $(\Yvect_n,\Xvect_n)$,
such that $\yvect_n$ 
is
the noisy image of $\xvect_n$ obtained from a $K$-component locally-affine transformation. 
This is modeled by introducing the latent variable $Z_n \in \{1,...,K \}$ such that: 
\begin{equation}
\label{eq:model_yxz}
  \Yvect_n=\sum_{k=1}^K \mathbb{I}(Z_n=k) (\Amat_k \Xvect_n + \bvect_k +
  \Evect_{k})
\end{equation}
where $\mathbb{I}$ is the indicator function, matrix
$\Amat_k\in\mathbb{R}^{D\times L}$ and vector
$\bvect_k\in\mathbb{R}^D$ define an affine transformation and
$\Evect_{k}\in\mathbb{R}^D$ is an error term capturing both the
observation noise in $\mathbb{R}^D$ and the reconstruction error
due to the local affine approximation. Under the assumption that
$\Evect_{k}$ is a zero-mean Gaussian variable with covariance
matrix $\Sigmamat_k\in\mathbb{R}^{D\times D}$ that does not
depend on $\Xvect_n$, $\Yvect_n$,  and $Z_n$, we obtain:
\begin{equation}
\label{eq:model_pYxz}
  p(\Yvect_n=\yvect_n|\Xvect_n=\xvect_n,Z_n=k;\thetavect) =
 \mathcal{N}(\yvect_n ;\Amat_k \xvect_n + \bvect_k ,\Sigmamat_k),
\end{equation}
where $\thetavect$ is the set of all the model parameters. It has to be noticed that the roles of the input and of the output (or response) variables are interchanged, \textit{i.e.}, the low-dimensional variable $\Xvect_n$ is the regressor.

To complete
the definition 
and enforce the affine
transformations to be local,  each $\Xvect_n$ is assumed to follow a mixture of $K$ Gaussians defined by
\begin{align}
 p(\Xvect_n=\xvect_n |Z_n=k; \thetavect) & = \mathcal{N}(\xvect_n ; \cvect_k,\Gammamat_k)
 \label{eq:Xz}
\end{align}
with means $\cvect_k\in\mathbb{R}^L$ and covariance matrices $\Gammamat_k\in\mathbb{R}^{L\times L}$. In the standard hGLLiM model, a multinomial prior
\begin{equation}
\label{eq:consprior}
 p(Z_n=k; \thetavect)  = \pi_k
\end{equation}
is used on the hidden variable $Z_n$. An efficient closed-form EM algorithm \cite{DeleforgeForbesHoraud2014}, provides maximum likelihood estimates of the parameters $\thetavect=\{\pi_k,\cvect_k,\Gammamat_k,\Amat_k, \bvect_k ,\Sigmamat_k\}$  given the observed training data $\{\yvect_n,\tvect_n\}_{n=1}^N$.

\subsection{Inverse Regression Strategy}
Once the parameter vector 
$\thetavect$ has been estimated, one obtains an inverse regression from $\mathbb{R}^L$ (low-dimensional space) to $\mathbb{R}^D$ (high-dimensional space), using 
the following \textit{inverse conditional density}:
\begin{align}
 & p(\Yvect_n=\yvect_n|\Xvect_n=\xvect_n;\thetavect)  = \nonumber \\
  \label{eq:JGMM_forward_map}
 & \sum_{k=1}^K \frac{p(Z_n=k)\mathcal{N}(\xvect_n ; \cvect_k,\Gammamat_k)}{\sum_{j=1}^Kp(Z_n=j)\mathcal{N}(\xvect_n ; \cvect_j,\Gammamat_j)} \mathcal{N}(\yvect_n;\Amat_k\xvect_n+\bvect_k,\Sigmamat_k).
\end{align}
The forward regression of interest, \textit{i.e.}, from $\mathbb{R}^D$ (the high dimension) to $\mathbb{R}^L$ (the low dimension),
is obtained from the \textit{forward conditional density}:
\begin{align}
& p(\Xvect_n=\xvect_n|\Yvect_n=\yvect_n;\thetavect) = \nonumber \\
 \label{eq:JGMM_inverse_map}
&\sum_{k=1}^K\frac{p(Z_n=k)\mathcal{N}(\yvect_n ; \cvect_k^*,\Gammamat_k^*)}{\sum_{j=1}^Kp(Z_n=j)\mathcal{N}(\yvect_n ; \cvect_j^*.\Gammamat^*_j)}
 \mathcal{N}(\xvect_n;\Amat^*_k\yvect_n+\bvect^*_k,\Sigmamat_k^*).
\end{align}
The latter involves the \textit{forward regression parameters}:
\begin{equation}
\label{eq:thetastar}
\thetavect^{*} = \{\cvect_k^{*},\Gammamat_k^{*}, \Amat_k^{*}, \bvect_k^{*},\Sigmamat_k^{*}\}_{k=1}^K
\end{equation}
that can be analytically derived from the \textit{inverse regression parameters} $\thetavect$ that were previously learned:
\begin{align}
\pi_k^* & = \pi_k, \\
  \cvect_k^* & =\Amat_k\cvect_k+\bvect_k, \label{eq:thetastarF} \\
  \Gammamat_k^* & =\Sigmamat_k+\Amat_k\Gammamat_k\Amat_k\tp, \\
  \Amat^*_k & = \Sigmamat_k^*\Amat_k\tp\Sigmamat_k^{-1},  \\
  \bvect^*_k & = \Sigmamat_k^*(\Gammamat_k^{-1}\cvect_k-\Amat_k\tp\Sigmamat_k^{-1}\bvect_k),\\
  \Sigmamat_k^* & = (\Gammamat_k^{-1}+\Amat_k\tp\Sigmamat_k^{-1}\Amat_k)^{-1}. \label{eq:thetastarL}
\end{align}
Learning an inverse regression model to obtain the desired forward mapping is referred to as \textit{inverse regression strategy}. This is motivated by a drastic reduction of the model's size, making tractable its estimation. Indeed, if we consider that the error vectors $\Evect_k$ are modeled by isotropic Gaussian noise with equal variances, the dimension of the learned parameter vector $\thetavect$ is $\mathcal{O}(DL+L^2)$, while it would be $\mathcal{O}(DL+D^2)$ using a forward model\footnote{Recall that $L\ll D$.}.

\subsection{Estimating the Latent Dimension with BIC}
\label{subsec:BIC}
The proposed model involves latent response variables $\Wvect_n$ of dimensionality $L_{\textrm{w}}$. The \textit{Bayesian information criterion} (BIC) can be used to select a value for the latent dimension.
Once a set of model parameters $\widetilde{\thetavect}$ has been learned, BIC can be computed as follows:
\begin{equation}
 BIC(\widetilde{\thetavect},N) = -2 \mathcal{L}(\widetilde{\thetavect}) + \mathcal{D}(\widetilde{\thetavect})\log N,
\end{equation}
where $\mathcal{L}$ denotes the observed-data log-likelihood and $\mathcal{D}(\widetilde{\thetavect})$ denotes the dimension of the complete parameter vector $\widetilde{\thetavect}$. Assuming  isotropic and equal noise covariance matrices $\{\Sigmamat_k\}_{k=1}^K$, we have:
\begin{align}
 \mathcal{D}(\widetilde{\thetavect}) &= K(D(L_{\textrm{w}}+L_{\textrm{t}}+1)+L_{\textrm{t}}(L_{\textrm{t}}+3)/2+1) \; , \\
 \mathcal{L}(\widetilde{\thetavect}) &= \sum_{n=1}^N\log p(\yvect_n,\tvect_n;\widetilde{\thetavect})\; ,
\end{align}
A natural way to define a value for $L_{\textrm{w}}$ is, for a given value of $K$, to train the model with different values of $L_{\textrm{w}}$, and select the value minimizing BIC. We will refer to the corresponding method as hGLLiM-BIC. While being computationally demanding, the BIC parameter selection procedure has the advantage of not requiring manually setting $L_\textrm{w}$. 

\section{Variational EM for the Spatial Markov Model}
\label{sec:variational-EM}
In this section we propose and derive an extension of hGLLiM that accounts for spatially located variables, \textit{i.e.}, $\Yvect_n$ is defined at the sites of hyper-spectral data, where each site $n \in S=\{1 \ldots N\}$ typically corresponds to a graph vertex or to a pixel. 
The novelty of the model proposed in this paper with respect to  \cite{DeleforgeForbesHoraud2014}   lies in the explicit spatial dependencies which are modeled by introducing the assumption that variables $\Zvect$ are distributed according to a discrete Markov random field (MRF) on $S$ with pairwise potentials. The multinomial prior (\ref{eq:consprior}) is therefore replaced by:
\begin{equation}
 p(\Zvect=\zvect; \psivect)  = {\cal K}(\psivect)^{-1}\;  \exp(H(\zvect; \psivect)) \label{def:mrf}
 \end{equation}
 with $H$ the energy function $H(\zvect; \psivect))= \sum_{n=1}^N \left( \alpha_{z_n} + \frac{\beta}{2} \sum_{m\in \Omega(n)} \mathbb{I}(z_n=z_m)\right) $ and the normalizing constant ${\cal K}(\psivect)= \sum\limits_{\zvect} \exp(H(\zvect; \psivect))$. 
In these expressions, $\Omega(n)$ denotes the neighbors of site $n$ in $S$ (typically the vertices linked to $n$ by an edge), $\psivect=\{\alphavect, \beta\}$, $\alphavect=(\alpha_1, \ldots, \alpha_K)$ is a vector of external field parameters and $\beta$ is a scalar interaction parameter. In practice, a constraint on $\alphavect$ is required to avoid identifiability issues. Thus, we assume $\alpha_1=0$. 
We also assume that variables from different site locations $\{(\Xvect_n, \Yvect_n)\}_{n=1}^N$ are conditionally independent given $\{Z_n\}_{n=1}^N$. 
Therefore, the parameters of this model are $\phivect=\{\thetavect, \psivect\}$ defined by:
\begin{eqnarray}
\label{eq:theta-def}
\thetavect &=& \{\cvect_k,\Gammamat_k,\Amat_k, \bvect_k,\Sigmamat_k\}_{k=1}^K \\
\psivect &=&\{\alphavect, \beta\}.  \label{eq:psi-def}
\end{eqnarray}

Because  $\Xvect_n$ defined in \eq{eq:partial-latent}
 is a \textit{partially-latent} variable, the parameter estimation procedure exploits observed pairs 
$\{\yvect_n,\tvect_n\}_{n=1}^N$ while being constrained by the presence of the latent variables $\Wvect_n$. 
The
decomposition of $\Xvect_n$  into observed and latent components implies that the model
parameters $\cvect_k$,
$\Gammamat_k$ and $\Amat_k$ must be decomposed as well. Assuming the independence of $\Tvect$
and $\Wvect$ given $\Zvect$, this gives:
\[
 \cvect_k=\left[
  \begin{array}{c}
   \cvect^{\textrm{t}}_{k}\\
   \cvect^{\textrm{w}}_{k}
  \end{array}
 \right],
 \Gammamat_k=\left[
  \begin{array}{lc}
    \Gammamat_k^{\textrm{t}}     & \zerovect \\
    \zerovect & \Gammamat_k^{\textrm{w}}
  \end{array}
 \right],
 \Amat_k= \left[ \begin{array}{cc} \Amat_k^{\textrm{t}} & \Amat_k^{\textrm{w}} \end{array} \right].
\]

Considering the complete data, with observed variables $(\Yvect_n,
\Tvect_n)_{1:N}$ and  hidden variables $(Z_n, \Wvect_n)_{1:N}$, the corresponding expectation-maximisation (EM) algorithm is intractable due to the complex Markovian dependencies between variables $Z_n$. Therefore, we resort to a 
 variational approximation which consists of approximating the
true posterior $p(\wvect, \zvect | \yvect, \tvect, \phivect)$ by a distribution  $q_{W,Z}(\wvect, \zvect)$ which factorizes  as 
$$ q_{W,Z}(\wvect, \zvect)= \prod\limits_{n=1}^N q_{W_n,Z_n}(\wvect_n, z_n).$$
From this factorization follows a variational EM (VEM)  parameter learning procedure which consists of $N$ successive E-steps denoted by E-$(W_n, Z_n)$ for $n=\{1,...,N\}$, followed by variational M-steps.

\subsection{Variational E-$(W_n, Z_n)$ step}
Given current  parameters values $\phivect^{(i)}$ and distribution $q_{W,Z}^{(i)}$,  the E-$(W_n, Z_n)$ step consists of
computing the probability distribution $q^{(i+1)}_{W_n,Z_n}(\wvect_n, z_n)$ which is proportional to:
\begin{equation}
\exp\left(\mathrm{E}_{q^{(i)}_{W\backslash n, Z\backslash n}}[ \; \log p(\wvect_n, z_n  \; | \; \yvect, \tvect, \Wvect\backslash n, \Zvect\backslash n ; \phivect^{(i)}) ]\right) \label{VEstep}
\end{equation}
where $\Wvect\backslash n$ (resp. $\Zvect\backslash n$) denotes the variables $\{\Wvect_m, m=1:N, m\not = n\}$ (resp. $\{Z_m, m=1:N, m\not = n\}$).

It follows from (\ref{VEstep})  that we can also write:
$$q^{(i+1)}_{W_n,Z_n}(\wvect_n, z_n)= q^{(i+1)}_{W_n | Z_n}(\wvect_n |  z_n) \; q^{(i+1)}_{Z_n}(z_n)$$
where:
\begin{align}
& q^{(i+1)}_{Z_n}(z_n=k) =  \nonumber \\
& \frac{p(\yvect_n, \tvect_n | z_n=k ; \thetavect^{(i)}) \; \exp(\alpha^{(i)}_k + \beta^{(i)} \sum_{m\in \Omega(n)} q^{(i)}_{Z_m}(k))}{\sum\limits_{l=1}^K p(\yvect_n, \tvect_n | z_n=l ; \thetavect^{(i)}) \; \exp(\alpha^{(i)}_l + \beta^{(i)} \sum_{m\in \Omega(n)} q^{(i)}_{Z_m}(l))} \qquad \\ 
& q^{(i+1)}_{W_n | Z_n}(\wvect_n |  z_n=k ) = {\cal N}(\wvect_n ; \widetilde{\muvect}_{nk}^{\textrm{w}}, \widetilde{\Smat}_k^{\textrm{w}})\label{def:qZ} 
\end{align}
with:
\begin{align}
\widetilde{\muvect}_{nk}^{\textrm{w}} & =  \widetilde{\Smat}_k^{\textrm{w}}\big((\Amat_k^{\textrm{w}(i)})^\top(\Sigmamat_k^{(i)})\inverse(\yvect_n-\Amat_k^{\textrm{t}(i)}\tvect_n-\bvect_k^{(i)}) \nonumber \\
&+ (\Gammamat_k^{\textrm{w}(i)})\inverse \cvect_k^{\textrm{w}(i)}\big) \\ 
\widetilde{\Smat}_k^{\textrm{w}}&= \big((\Gammamat_k^{\textrm{w}(i)})\inverse +(\Amat_k^{\textrm{w}(i)})^\top(\Sigmamat_k^{(i)})\inverse \Amat_k^{\textrm{w}(i)}\big)\inverse .
\end{align}

\subsection{\addnote[vmstep]{1}{Variational M-Steps}}
\label{subsec:variational}
The variational M-steps consist in estimating the parameters $\phivect=\{\thetavect,\psivect\}$ that maximize the expectation:
\begin{equation}
\mathrm{E}_{q^{(i+1)}_{W,Z}}[\log p(\yvect,\tvect, \Wvect, \Zvect ;  \thetavect, \psivect)]. \label{def:M-step}
\end{equation}
It follows that the update of parameters in $\thetavect$ is identical to that of the standard hGLLiM EM \cite{DeleforgeForbesHoraud2014}, where the posterior densities $p(Z_n=k | \yvect_n,\tvect_n ;\thetavect^{(i)})$ are now replaced by the variational distributions $q^{(i+1)}_{Z_n}(z_n=k)$. As in \cite{DeleforgeForbesHoraud2014}, the update of noise covariance matrices $\{\Sigmamat_k\}_{k=1}^K$ depends on the way they are constrained. In practice, we enforce isotropic covariances for all $k$ in order to avoid over-parameterization.

Once $\thetavect$ has been updated, the proposed MRF model induces an additional M-step for the update of $\psivect=\{\alphavect, \beta\}$.
This step does not admit an explicit closed-form expression but can be solved numerically using  gradient descent schemes.  It is straightforward to show that the maximization of (\ref{def:M-step}) in $\psivect$ admits a unique solution. Indeed, it is equivalent to solve
\begin{eqnarray}
\psivect^{(i+1)}&=& \arg\max_{\psivect}\mathrm{E}_{q^{(i+1)}_{Z}}[\log p(\Zvect ;  \psivect)]\nonumber\\
&=&  \arg\max_{\psivect}\mathrm{E}_{q^{(i+1)}_{Z}}[H(\zvect; \psivect)] - \log {\cal K}(\psivect) \label{def:M-psi} \;.
\end{eqnarray}
Denoting the gradient vector and Hessian matrix respectively by $\nabla_{\psivect}$ and $\nabla^2_{\psivect}$, it comes
\begin{eqnarray}
\nabla_{\psivect}\mathrm{E}_{q^{(i+1)}_{Z}}[\log p(\Zvect ;  \psivect)] &=& \mathrm{E}_{q^{(i+1)}_{Z}}[\nabla_{\psivect}H(\Zvect; \psivect)] \nonumber \\
&-& \mathrm{E}_{p(z ; \psivect)}[\nabla_{\psivect}H(\Zvect; \psivect)] \label{grad}\\
\nabla^2_{\psivect} \mathrm{E}_{q^{(i+1)}_{Z}}[\log p(\Zvect ;  \psivect)]&=&  \mathrm{E}_{q^{(i+1)}_{Z}}[\nabla^2_{\psivect}H(\Zvect; \psivect)] \nonumber \\
& -& \mathrm{E}_{p(z ; \psivect)}[\nabla^2_{\psivect}H(\Zvect; \psivect)] \nonumber \\
& - & \mathrm{var}_{p(z ; \psivect)}[\nabla_{\psivect}H(\Zvect; \psivect)] \label{hess} \;.
\end{eqnarray}
The last expectations in (\ref{grad}) and (\ref{hess}) are taken over the Potts prior (\ref{def:mrf}). It follows that whenever $H(\zvect ; \psivect)$ is linear in $\psivect$, $\nabla^2_{\psivect}H(\Zvect; \psivect)$ is zero, the Hessian matrix is a semi-definite negative matrix and the function to optimize is concave. Unfortunately, due to the intractable normalizing constant ${\cal K}$, expressions (\ref{grad}) and (\ref{hess}) are not directly available. It is necessary to approximate the terms involving the true MRF prior $p(\zvect ; \psivect)$ 
using  an approximation. A natural approach is to use:
 \begin{equation}
 q^{prior}_Z(\zvect; \alphavect, \beta) = \prod\limits_{n=1}^N q^{prior}_{Z_n}(z_n ; \alphavect, \beta)
\end{equation}
with $q^{prior}_{Z_n}(z_n; \alphavect, \beta)$ defined by:  
\begin{eqnarray}
q^{prior}_{Z_n}(z_n=k; \alphavect, \beta) = \frac{\exp(\alpha_k + \beta \sum_{m\in \Omega(n)} q^{(i)}_{Z_m}(k))}{\sum\limits_{l=1}^K  \; \exp(\alpha_l + \beta \sum_{m\in \Omega(n)} q^{(i)}_{Z_m}(l))}\; \label{def:pikn}
\end{eqnarray}
 This MRF prior approximation induced by the posterior variational approximation has been proposed in \cite{Celeux2003} and also exploited in \cite{Chaari13}. 
%
It follows that the gradient in (\ref{grad}) can be approximated by (with $\alpha_1=0$):
\begin{equation}
\left\{
\begin{array}{l}
\frac{1}{2}  \sum\limits_{n=1}^N   \sum\limits_{m\in \Omega(n)} \sum\limits_{k=1}^K \big( q_{Z_n}^{(i+1)}(k)  \; q_{Z_m}^{(i+1)}(k)  \\
\hspace{10pt} - q_{Z_n}^{prior}(k; \alphavect, \beta)  \; q_{Z_m}^{prior}(k; \alphavect, \beta) \big)   \\
\sum\limits_{n=1}^N  \big( q_{Z_n}^{(i+1)}(2)   - q_{Z_n}^{prior}(2; \alphavect, \beta)   \big) \\
\vdots \\
\sum\limits_{n=1}^N  \big( q_{Z_n}^{(i+1)}(K)   - q_{Z_n}^{prior}(K; \alphavect, \beta)  \big) 
\end{array}
\right. 
\end{equation}

\subsection{Inverse Regression Strategy with Markov Random Fields}
\label{subsec:inverseMRF}
With this proposed MRF extension of hGLLiM, the derivation of the desired forward conditional density (\ref{eq:JGMM_inverse_map}) is less straightforward. Indeed, while equations (\ref{eq:JGMM_inverse_map})-(\ref{eq:thetastar}) remain valid, they now involve the marginal of the MRF prior $p(Z_n=k)$ which is known to be hard to compute (see (\ref{def:mrf})). Approximations are required and a natural 
candidate is $q_{Z_n}^{prior}(k)$ which has to be computed iteratively using the variational EM algorithm with $\thetavect$  (or $\thetavect^{*}$) held fixed. Note that $\alphavect$ could also either be set fixed to the learned values or re-estimated. At iteration (i), for each $n$,  given current values for $\alphavect^{(i)}, \beta^{(i)}$ and $q^{(i)}_{Z_m}(k)$, $q_{Z_n}(k)$ is updated using:
\begin{align}
&q^{(i+1)}_{Z_n}(z_n=k) \nonumber \\
&= \frac{p(\yvect_n | z_n=k ; \thetavect^{*}) \; \exp(\alpha^{(i)}_k + \beta^{(i)} \sum_{m\in \Omega(n)} q^{(i)}_{Z_m}(k))}{\sum\limits_{l=1}^K p(\yvect_n | z_n=l ; \thetavect^{*}) \; \exp(\alpha^{(i)}_l + \beta^{(i)} \sum_{m\in \Omega(n)} q^{(i)}_{Z_m}(l))}. \qquad \label{def:qZ2}
\end{align}
Although the update formula for $q_{Z_n}(k)$ is synchronous, an asynchronous update is known to be more efficient. Then parameters  $\alphavect^{(i+1)}, \beta^{(i+1)}$ are obtained by solving (\ref{def:M-psi}) as explained in Section \ref{subsec:variational}. 
At algorithm convergence, $p(Z_n = k)$ is set as:
\begin{align}
p(Z_n = k) & \approx q_{Z_n}^{prior}(k; \alphavect^{(\infty)}, \beta^{(\infty)}) \nonumber \\
\label{eq:proba_Zn_equal_k}
& = \frac{\exp(\alpha^{(\infty)}_k + \beta^{(\infty)} \sum_{m\in \Omega(n)} q_{Z_m}^{(\infty)}(k))}{\sum_{l=1}^K \exp(\alpha^{(\infty)}_l+ \beta^{(\infty)} \sum_{m\in \Omega(n)} q_{Z_m}^{(\infty)}(l))}.
\end{align}
%
When required, a response $\xvect_n$ corresponding to an input $\yvect_n$ can be obtained using the expectation of $p(\Xvect_n=\xvect_n|\Yvect_n=\yvect_n)$ that can be approximated using:
\begin{align}
& \mathrm{E}[\Xvect_n|\yvect_n;\thetavect^*, \psivect] \approx \nonumber \\  
\label{eq:JGMM_inverse_exp_MRF}
& \sum_{k=1}^K \frac{q_{Z_n}^{prior}(k; \alphavect^{(\infty)}, \beta^{(\infty)}) \mathcal{N}(\yvect_n; \cvect_k^*,\Gammamat_k^*)}{\sum_{j=1}^K q_{Z_n}^{prior}(j; \alphavect^{(\infty)}, \beta^{(\infty)})\mathcal{N}(\yvect_n ; \cvect_j^*,\Gammamat^*_j)}(\Amat^*_k\yvect_n+\bvect^*_k).
\end{align}


%



\subsection{\addnote[outline]{1}{Prediction with spatial constraints: MRF-hGLLiM}}
To complete this section, we provide an overview of the proposed procedure implementing the inverse regression strategy while accounting for spatial interactions. It divides into two steps, a learning step followed by a prediction step. Each step is summarized below:

\begin{itemize}

\item[] \textbf{Spatial inverse-regression learning:}
\begin{itemize}
\item {\it Training observed data:}   $\{(\yvect_n, \tvect_n)\}_{n=1}^N $;
\item {\it Missing variables:}  $\{(\Wvect_n, \Zvect_n)\}_{n=1}^N$;
\item {\it Model: MRF-hGLLiM}  (hybrid Gaussian locally linear mapping with spatial constraints) defined by \eqref{eq:partial-latent}, \eqref{eq:model_pYxz}, \ref{eq:Xz}, and \eqref{def:mrf});
\item \textit{Parameters:}  $\thetavect = \{\cvect_k,\Gammamat_k,\Amat_k, \bvect_k,\Sigmamat_k\}_{k=1}^K $ and $\psivect =\{\alphavect, \beta\}$;
\item {\it Estimation procedure:} VEM (section \ref{sec:variational-EM});
\item {\it Output:} 
\begin{itemize}
\item[-] $\hat{\thetavect}$ : estimated inverse regression parameters $\thetavect$,
\item[-] $\hat{\thetavect}^*$ : estimated forward regression parameters $\thetavect^*$, defined by (\ref{eq:thetastarF}) to (\ref{eq:thetastarL}),
\item[-] $\hat{\psivect}$ : estimated Markov parameters $\psivect$.
\end{itemize}
\end{itemize}
\item[] \textbf{Spatial forward-prediction:}
\begin{itemize}
\item {\it Observed data:}   $\{\tilde{\yvect}_n\}_{n=1}^{\tilde{N}}$ (no observed  $\{\tilde{\tvect}_n\}_{n=1}^{\tilde{N}} $);
\item {\it Missing variables:} $\{(\Xvect_n, \Zvect_n)\}_{n=1}^{\tilde{N}}$  where $\Xvect_n = [\Tvect_n;\Wvect_n]$; 
\item {\it Model:}  MRF-hGLLiM with $\thetavect$ fixed to $\hat{\thetavect}$;
\item  {\it Estimation procedure:} VEM (section \ref{sec:variational-EM}) with no M-$\thetavect$ step ($\thetavect$ fixed to $\hat{\thetavect}$) which reduces to the procedure described in section \ref{subsec:inverseMRF};
\item {\it Output:}
\begin{itemize}
\item[-] $\tilde{\psivect} = \{\tilde{\alphavect}, \tilde{\beta}\}$,
\item[-] Estimated predictions for $\{\xvect_n\}_{n=1}^{\tilde{N}}$, \textit{i.e.}, $\tilde{\xvect}_n = \mathrm{E}[\Xvect_n|\tilde{\yvect}_n; \hat{\thetavect}^*, \tilde{\psivect}]$ computed using approximation (\ref{eq:JGMM_inverse_exp_MRF}).
\end{itemize}
\end{itemize}
\end{itemize}

\section{Retrieving Mars Physical Properties from Hyper-spectral Images}
\label{sec:applications}
In order to evaluate the proposed model, we use a database composed of synthetic spectra with their associated physical
parameter values. This database was generated using the radiative transfer
model presented in \cite{Doute2007} in order to investigate hyperspectral images
collected from the imaging spectrometer OMEGA \cite{Bibring2004} onboard of the Mars Express spacecraft. \addnote[physical3]{1}{The synthetic dataset is composed of 15,407 spectra associated with five
 physical parameter values, namely:}

\begin{itemize}
\item[(i)] {\em \addnote[physical4]{1}{proportion of water
ice}} (Prop. H$_2$O), 
\item[(ii)] {\em proportion of CO$_2$ ice} (Prop. CO$_2$), 
\item[(iii)] {\em proportion of
dust} (Prop. Dust), 
\item[(iv)] {\em grain size of water ice} (Size H$_2$O), and 
\item[(v)] {\em grain size of CO$_2$ ice} (Size CO$_2$). 
\end{itemize}
Each spectrum is made of 184 wavelengths. \addnote[otherdata]{1}{Compared to other available hyper-spectral image databases, \textit{e.g.} \cite{leenaars2013africa,tits2014validating}, this one combines several desirable features, making it suitable for the evaluation of the proposed model for the following reasons. First, the radiative transfer model can be used to generate as much data as required to reliably estimate the model parameters. Second, generated spectra present a highly non-linear relationship with underlying physical parameters. Third, the associated real hyper-spectral images of Mars surface contain spatial dependencies between neighboring pixels.}

\subsection{Hybrid-GLLiM Evaluation on Synthetic Individual Spectra}
In order to numerically evaluate the performances of MRF-hGLLiM on the synthetic data, we adopted a cross-validation strategy: one part of the
synthetic database is used for training and the remaining part is used for testing.
Note that because all these spectra were generated independently, there are no spatial dependencies present in the training dataset. Therefore, the value of the MRF interaction parameter $\beta$ was set to 0 in that case, \textit{i.e.}, no neighborhood dependencies are accounted for during training. With this setting, the proposed MRF-hGLLiM and hGLLiM models are equivalent.
The synthetic data are used, firstly to learn an inverse \textit{low-dimensional to high-dimensional regression function} between
physical parameters and spectra from the database, and secondly to estimate
the unknown physical parameters corresponding to a newly observed individual spectrum and using the forward mapping (\ref{eq:JGMM_inverse_exp_MRF}), where $q_{Z_n}^{prior}(k)$ is replaced by $e^{\alpha_k}$, or equivalently $\pi_k$.

\begin{table*}
\caption{\label{tab:planeto_full_err} Normalized root mean squared error (NRMSE) for Mars surface physical properties recovered from
individual hyper-spectral vectors, using synthetic data, different methods and fully-observed-output training.}
   \centering
   \begin{tabular}{|c|c|c|c|c|c|}
       \hline
       Method & Prop. H$_2$O  & Prop. CO$_2$  & Prop. Dust  & Size H$_2$O   & Size CO$_2$\\
       \hline
       JGMM   & \footnotesize{$2.40\pm18.5$} & \footnotesize{$0.84\pm1.64$} & \footnotesize{$0.63\pm1.02$} & \footnotesize{$0.73\pm1.02$} & \footnotesize{$1.08\pm4.52$} \\
       \hline
       SIR-1  & \footnotesize{$3.41\pm20.0$} & \footnotesize{$1.28\pm2.16$} & \footnotesize{$1.04\pm1.79$} & \footnotesize{$0.69\pm0.92$} & \footnotesize{$1.85\pm7.24$} \\
       \hline
       SIR-2  & \footnotesize{$3.27\pm18.6$} & \footnotesize{$0.96\pm1.75$} & \footnotesize{$0.89\pm1.53$} & \footnotesize{$0.62\pm0.86$} & \footnotesize{$1.66\pm6.53$} \\
       \hline
       RVM    & \footnotesize{$1.28\pm7.57$} & \footnotesize{$0.50\pm0.95$} & \footnotesize{$0.40\pm0.69$} & \footnotesize{$0.51\pm0.67$} & \footnotesize{$0.89\pm3.80$} \\
       \hline
       \textbf{MLE (hGLLiM-0)}   & {\boldmath \footnotesize{$1.04\pm6.66$}} & {\boldmath \footnotesize{$0.37\pm0.72$}} & {\boldmath \footnotesize{$0.28\pm0.50$}} & {\boldmath \footnotesize{$0.45\pm0.74$}} & {\boldmath \footnotesize{$0.60\pm2.59$}} \\
       \hline
       \textbf{hGLLiM-1} & {\boldmath \footnotesize $0.95\pm5.92$} & {\boldmath \footnotesize $0.34\pm0.65$} & {\boldmath \footnotesize $0.24\pm0.44$} & {\boldmath \footnotesize  $0.42\pm0.71$} & {\boldmath \footnotesize $0.56\pm2.44$} \\
       \hline
       \textbf{hGLLiM-2} & {\boldmath \footnotesize{$0.99\pm6.02$}} & {\boldmath \footnotesize{$0.36\pm0.70$}} & {\boldmath \footnotesize{$0.27\pm0.48$}} & {\boldmath \footnotesize{$0.40\pm0.66$}} & {\boldmath \footnotesize{$0.58\pm2.66$}} \\
       \hline
\end{tabular}
\end{table*}

\begin{table*}
\caption{\label{tab:planeto_err} Normalized root mean squared error (NRMSE) for Mars surface physical properties recovered from
individual hyper-spectral vectors, using synthetic data, different methods and partially-observed-output training.}
   \centering
   \begin{tabular}{|c|c|c|c|}
       \hline
       Method & Proportion of CO$_2$ ice & Proportion of dust & Grain size of H$_2$O ice \\
       \hline
       JGMM   & $0.83\pm1.61$ & $0.62\pm1.00$ & $0.79\pm1.09$  \\
       \hline
       SIR-1  & $1.27\pm2.09$ & $1.03\pm1.71$ & $0.70\pm0.94$  \\
       \hline
       SIR-2  & $0.96\pm1.72$ & $0.87\pm1.45$ & $0.63\pm0.88$  \\
       \hline
       RVM    & $0.52\pm0.99$ & $0.40\pm0.64$ & $0.48\pm0.64$ \\
       \hline
       MLE (hGLLiM-0)    & $0.54\pm1.00$ & $0.42\pm0.70$ & $0.61\pm0.92$ \\
       \hline
       hGLLiM-1 & $0.36\pm0.70$ & $0.28\pm0.49$ & $0.45\pm0.75$  \\
       \hline
       \textbf{hGLLiM-2}{\boldmath ${}^{*\dagger}$} &  {\boldmath $0.34\pm0.63$} & {\boldmath $0.25\pm0.44$} & {\boldmath $0.39\pm0.71$} \\
       \hline
       hGLLiM-3 &  $0.35\pm0.66$ & $0.25\pm0.44$ & $0.39\pm0.66$ \\
       \hline
       hGLLiM-4 & $0.38\pm0.71$ & $0.28\pm0.49$ & $0.38\pm0.65$ \\
       \hline
       hGLLiM-5 & $0.43\pm0.81$ & $0.32\pm0.56$ & $0.41\pm0.67$ \\
       \hline
       hGLLiM-20 & $0.51\pm0.94$ & $0.38\pm0.65$ & $0.47\pm0.71$ \\
       \hline
       \textbf{hGLLiM-BIC} &  {\boldmath $0.34\pm0.63$} & {\boldmath $0.25\pm0.44$} & {\boldmath $0.39\pm0.71$} \\
       \hline
\end{tabular}
\end{table*}

The hGLLiM algorithm was compared to JGMM
(joint Gaussian mixture model) \cite{QiaoMinematsu09}, MLE (mixture of linear experts) \cite{XuJordanHinton95}, SIR (sliced inverse regression) \cite{Li91} and RVM (multivariate relevance vector machine) \cite{thayananthan2006multivariate}.
When a value $L_{\textrm{w}}>0$ is used for training, we denote by hGLLiM-$L_{\textrm{w}}$ the hybrid GLLiM algorithm.
As shown in \cite{DeleforgeForbesHoraud2014}, JGMM and MLE are equivalent to hGLLiM-0 with covariance matrices $\{\Sigmamat_k\}_{k=1}^K$ being respectively unconstrained and diagonal. 
SIR is used with one (SIR-1) or two
(SIR-2) principal axes for dimensionality reduction, 20 slices\footnote{The number of slices is known to have very little influence on
the results.}, and polynomial regression of order three\footnote{Higher
orders did not show significant improvements in our experiments.}. SIR quantizes the
low-dimensional data $\Xvect$ into {\it slices} or clusters which
in turn induces a quantization of the $\Yvect$-space. Each
$\Yvect$-slice (all points $\yvect_n$ that map to the same
$\Xvect$-slice) is then replaced with its mean and PCA is carried
out on these means. The resulting dimensionality reduction is then
informed by $\Xvect$ values through the preliminary slicing. RVM may be viewed as a multivariate probabilistic formulation of \textit{support vector regression} \cite{smola2004tutorial}.
As with any kernel method, RVM critically depends on the choice of a kernel function. Using the authors'
publicly available software\footnote{http://www.mvrvm.com/Multivariate\_Relevance\_Vector\ },
we ran preliminary tests to determine an optimal kernel choice for each
dataset under consideration. We tested 14 kernels with 10 different scales, ranging from 1 to
30, resulting in a total of 140 kernels for each dataset.

As already mentioned an objective evaluation is performed
using cross-validation. A total of 10,000 training input-output couples are randomly selected twenty times as the training set, and the 5,407 remaining
spectra are used for testing.
For all models, the training data are normalized to zero mean and unit variance. Normalization is then reversed on test data and estimated output to obtain the final results. This normalization showed to significantly improve the performances of all methods. We use $K=50$ for MLE, hGLLiM and JGMM. MLE and JGMM are constrained with isotropic covariance matrices as this parametrization achieves the best results. For RVM, a third degree polynomial kernel with scale
6 is selected as the top performing among 140 tested
kernels using cross-validation. 

As a quality measure of the
estimated parameters, we compute the normalized root mean-square
error (NRMSE) which quantifies
the difference between estimated and ground-truth parameter values $\hat{t}_m$ and $t_m$:
\begin{equation}
 \mbox{NRMSE}=\sqrt{\frac{\sum_{m=1}^M (\hat{t}_m - t_m)^2}{\sum_{m=1}^M (t_m - \overline{t})^2}} \mbox{  with  } \overline{t} = \frac{1}{M}\sum_{m=1}^M t_m.
\end{equation}
NRMSE is normalized thus enabling direct comparison between the
parameters which are of very different range: closer
to zero, more accurate the predicted values. 

\paragraph{Fully Observed Output}
\label{par:fully_obs}
Using these different regression models, from 184-dimensional spectra, we start by retrieving the five parameter values, namely, proportion of water
ice (Prop. H$_2$O), proportion of CO$_2$ ice (Prop. CO$_2$), proportion of
dust (Prop. Dust), grain size of water ice (Size H$_2$O), and grain size of CO$_2$ ice (Size CO$_2$).
The training is done with synthetic spectra annotated with the five parameters, and hence output variables are fully observed. Results obtained with the five methods are summarized in Table \ref{tab:planeto_full_err}.
MLE and hGLLiM perform similarly and outperform the three other methods in estimating each parameter. \addnote[complexity1a]{1}{The average training time of MLE was 12.6 seconds and the average training time of hGLLiM was 18.9 seconds using our Matlab implementation}.
In this task, as expected, using 1 or 2 additional latent components in hGLLiM
only shows a slight improvement compared to MLE, since the output is fully observed during training. Notice that the obtained mean NRMSE with the proportion of water (Table \ref{tab:planeto_full_err}, column 2) and the grain size of CO$_2$ (Table \ref{tab:planeto_full_err}, column 6) parameters are very large, \textit{i.e.}, above $0.5$ for all methods. This suggests that the relationship between these parameters and observed spectra is complex and hard to learn.

\begin{figure*}
     \centering
     \subfigure[Ground-truth]{
       \includegraphics[width=.26\linewidth,clip=,keepaspectratio]{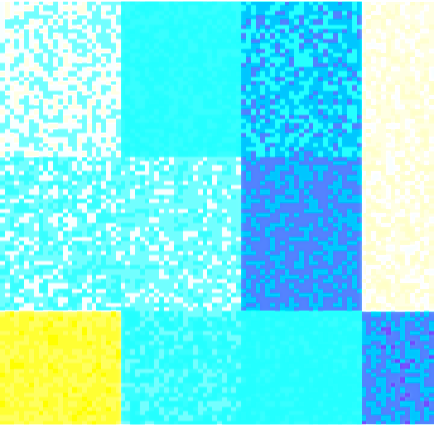}
             \hfill }
        \subfigure[Estimated with MRF-hGLLiM-0]{
      \includegraphics[width=.26\linewidth,clip=,keepaspectratio]{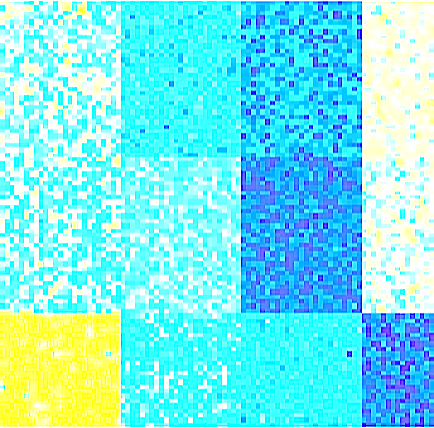}
      \hfill  
    }   
    \subfigure[Estimated with hGLLiM-0]{
      \includegraphics[width=.26\linewidth,clip=,keepaspectratio]{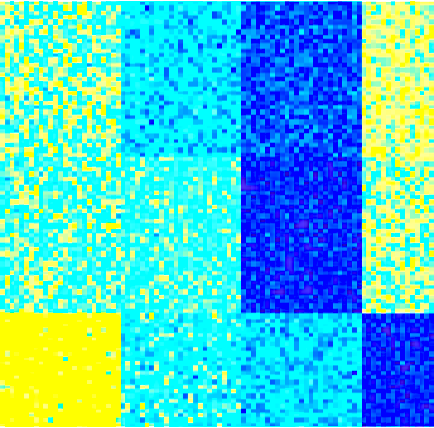}
      \hfill  
    }
     \caption{\label{fig:synth_spatial}Ground truth and estimated dust concentration maps with  MRF-hGLLiM  and hGLLiM using synthetic hyper-spectral images. For visualization purpose, the image color dynamics have been adjusted to emphasize the differences.} 
 \end{figure*}

\paragraph{Partially Latent Output}
In order to fully illustrate the
potential of hybrid GLLiM, we deliberately ignore the values of two of the five parameters in
the database and treat these two parameters as \textit{latent variables}.  
We chose to ignore the proportion of water ice and the grain size of CO$_2$
ice since they yield the worst reconstruction errors when they are fully observed (see Table \ref{tab:planeto_full_err}).
A previous study \cite{bernard2009retrieval} reveals that these two parameters are sensitive to the same
wavelengths and are suspected to \textit{mix}
with the other three parameters in the radiative transfer model used to generate the synthetic data. Therefore, not only that they are difficult to estimate, but they also seem to affect the estimation of the three other parameters.

\begin{table*}
\caption{\label{tab:planeto_spatial_err} Normalized root mean squared error (NRMSE) of hGLLiM-0 and MRF-GLLiM for Mars surface physical properties recovered from
hyper-spectral images, using synthetic data and fully-observed-output training.}
   \centering
   \begin{tabular}{|c|c|c|c|c|c|}
       \hline
       Method & Prop. H$_2$O  & Prop. CO$_2$  & Prop. Dust  & Size H$_2$O   & Size CO$_2$\\
       \hline
       hGLLiM-0   & \footnotesize{$0.27\pm0.02$} & \footnotesize{$0.25\pm0.03$} & \footnotesize{$0.25\pm0.03$} & \footnotesize{$0.29\pm0.05$} & \footnotesize{$0.24\pm0.04$} \\
       \hline
       \textbf{MRF-hGLLiM-0} & {\boldmath \footnotesize{$0.25\pm0.03$}} & {\boldmath \footnotesize{$0.23\pm0.03$}} & {\boldmath \footnotesize{$0.23\pm0.03$}} & {\boldmath \footnotesize{$0.26\pm0.05$}} & {\boldmath \footnotesize{$0.23\pm0.04$}} \\
       \hline
\end{tabular}
\end{table*}

\begin{figure*}[p!]
    \centering
    \subfigure[Proportion of dust]{
    \begin{minipage}[c]{.200\linewidth}
    \centering
    \includegraphics[width=.95\linewidth,clip=,keepaspectratio]{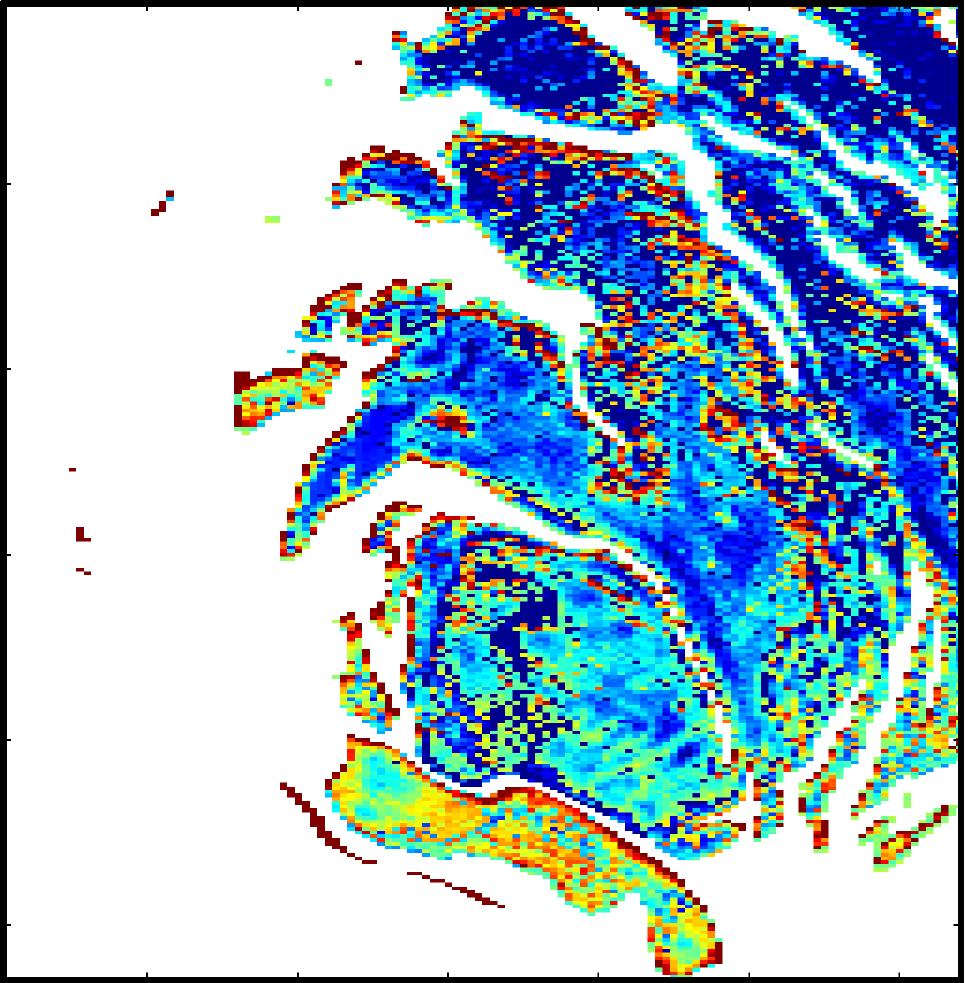}
    \includegraphics[width=.95\linewidth,clip=,keepaspectratio]{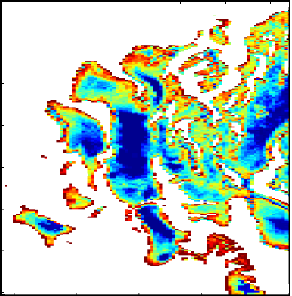}
    \\ hGLLiM-2
   \end{minipage} \hfill
   \begin{minipage}[c]{.200\linewidth}
    \centering
    \includegraphics[width=.95\linewidth,clip=,keepaspectratio]{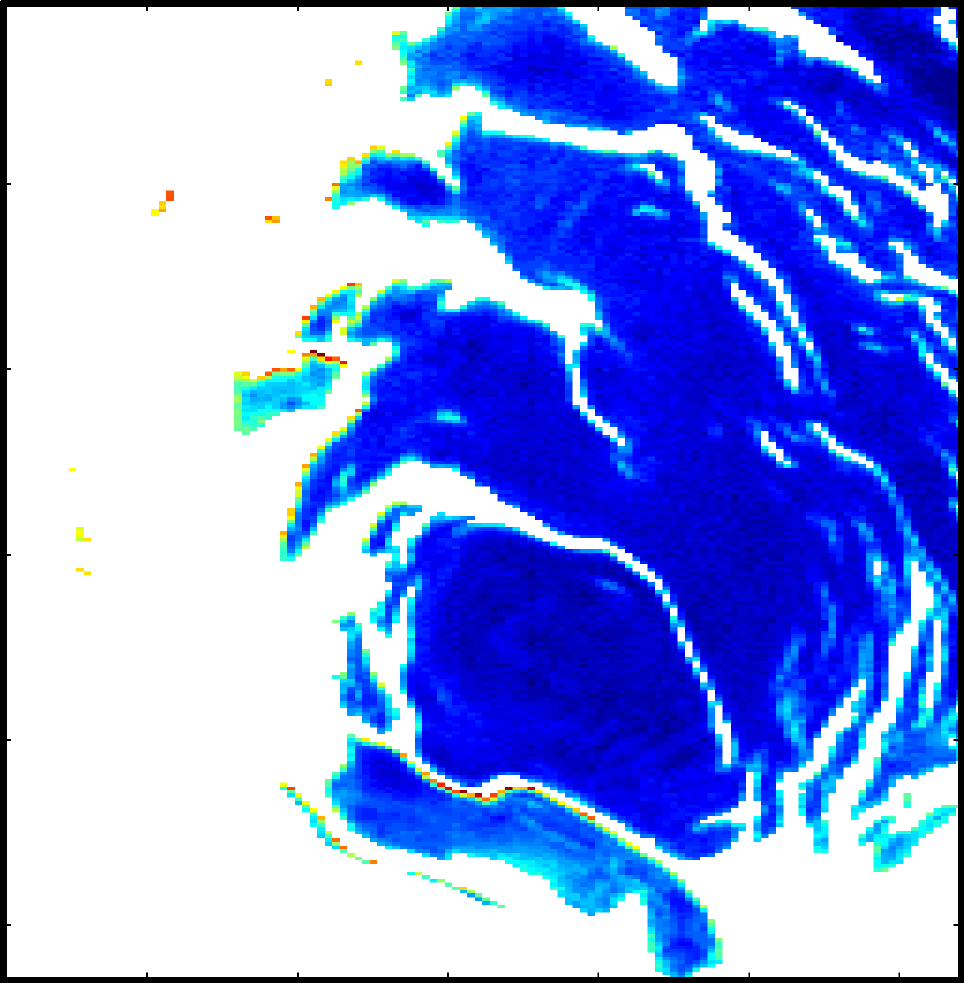}
    \includegraphics[width=.95\linewidth,clip=,keepaspectratio]{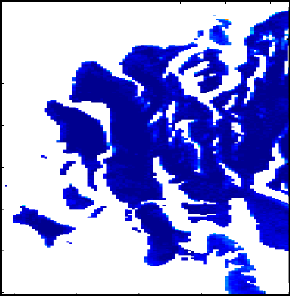}
    \\ RVM
   \end{minipage} \hfill
   \begin{minipage}[c]{.200\linewidth}
    \centering
    \includegraphics[width=.95\linewidth,clip=,keepaspectratio]{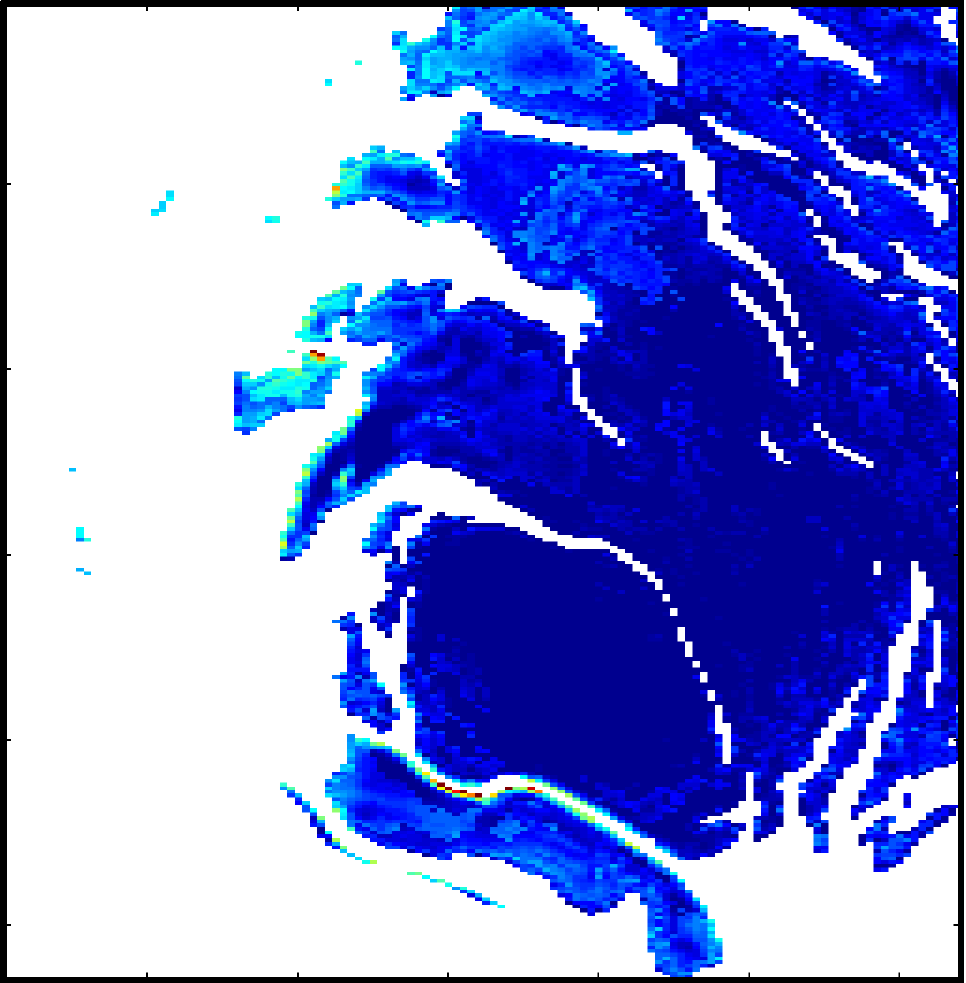}
    \includegraphics[width=.95\linewidth,clip=,keepaspectratio]{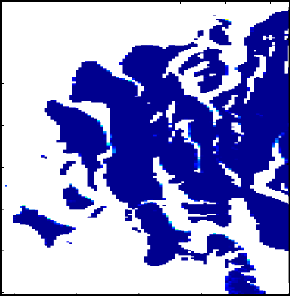}
    \\ MLE
   \end{minipage} \hfill
   \begin{minipage}[c]{.200\linewidth}
    \centering
    \includegraphics[width=.95\linewidth,clip=,keepaspectratio]{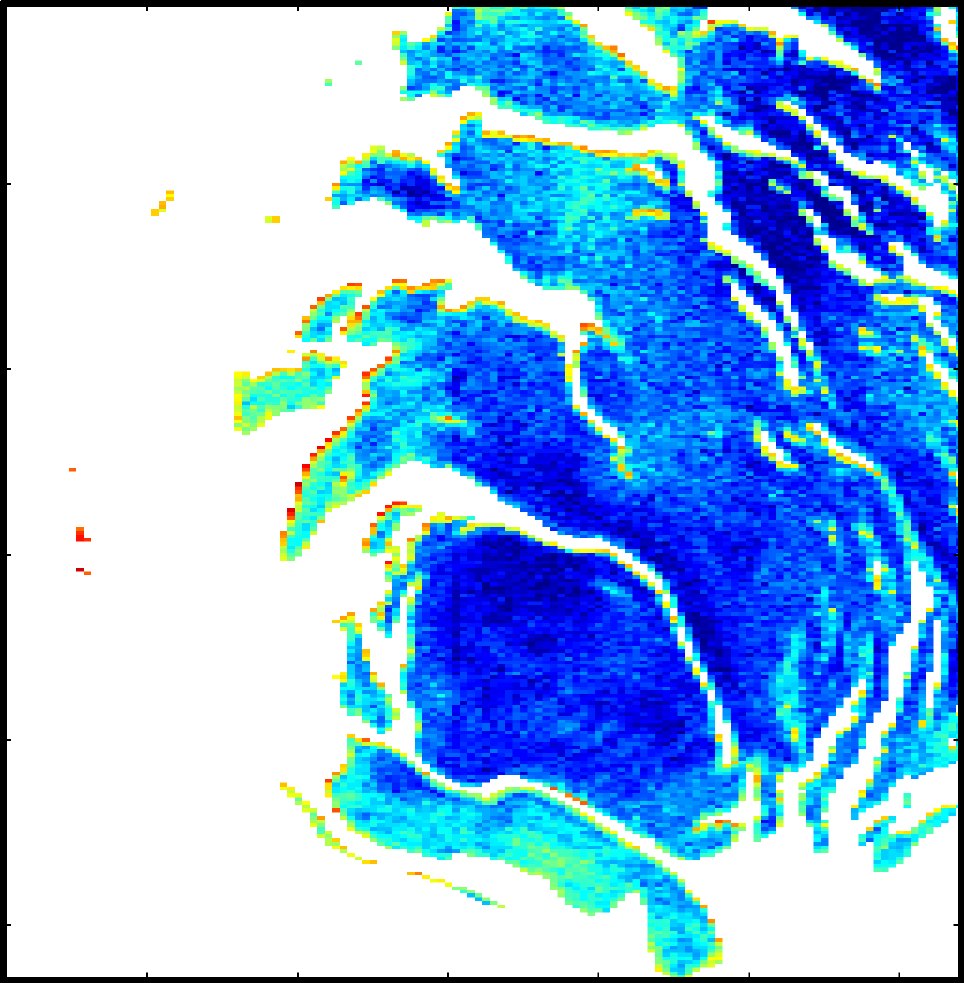}
    \includegraphics[width=.95\linewidth,clip=,keepaspectratio]{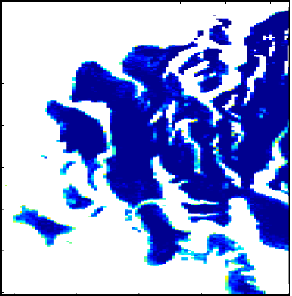}
    \\ JGMM
   \end{minipage} \hfill
   \begin{minipage}[c]{.061\linewidth}
    \centering
    \includegraphics[width=.95\linewidth,clip=,keepaspectratio]{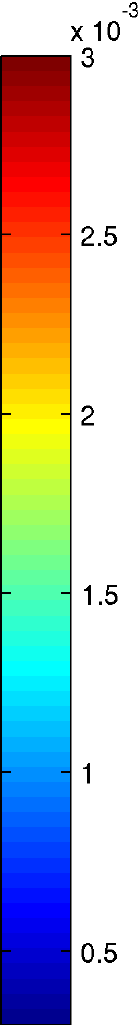}
    \\ $\mbox{ }$
   \end{minipage} \hfill
   }
   \subfigure[Proportion of CO$_2$ ice]{
   \begin{minipage}[c]{.200\linewidth}
    \centering
    \includegraphics[width=.95\linewidth,clip=,keepaspectratio]{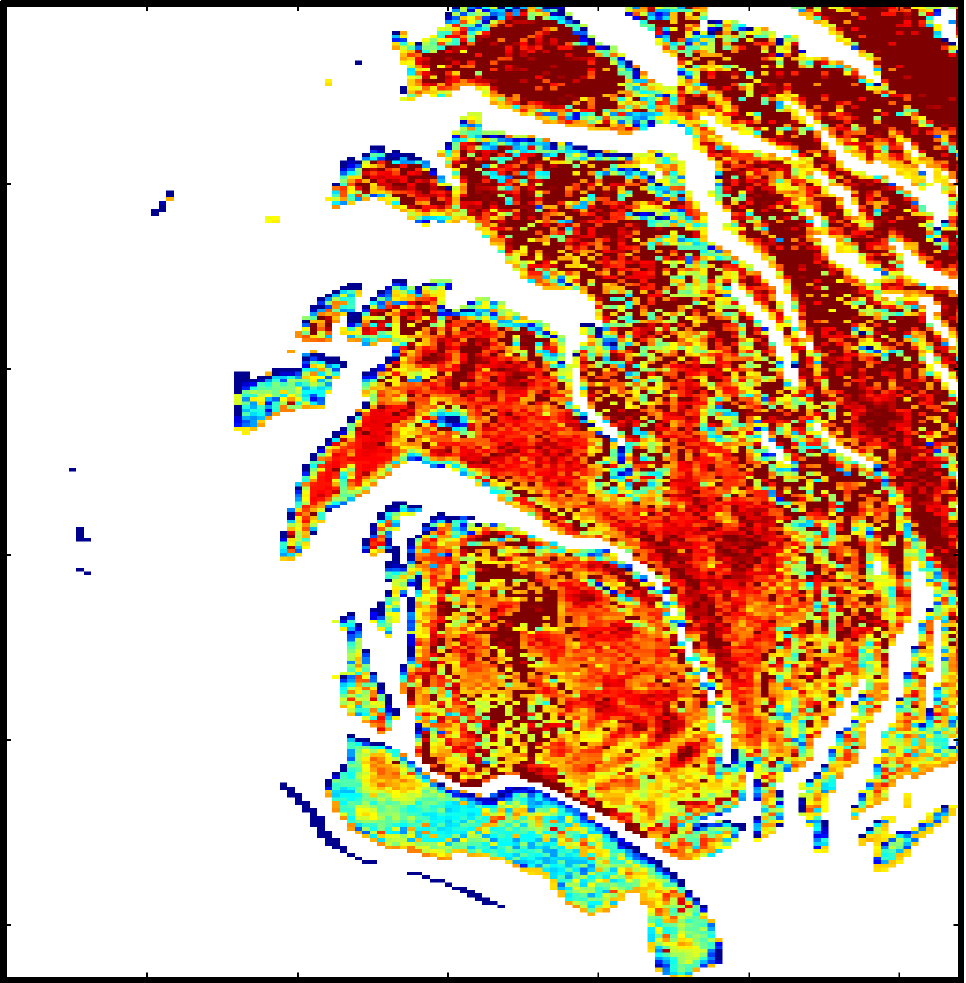}
    \includegraphics[width=.95\linewidth,clip=,keepaspectratio]{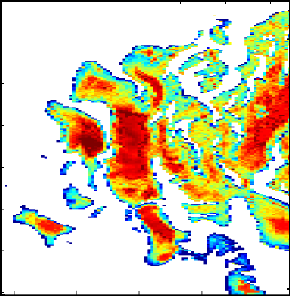}
     \\ hGLLiM-2
   \end{minipage} \hfill
   \begin{minipage}[c]{.200\linewidth}
    \centering
    \includegraphics[width=.95\linewidth,clip=,keepaspectratio]{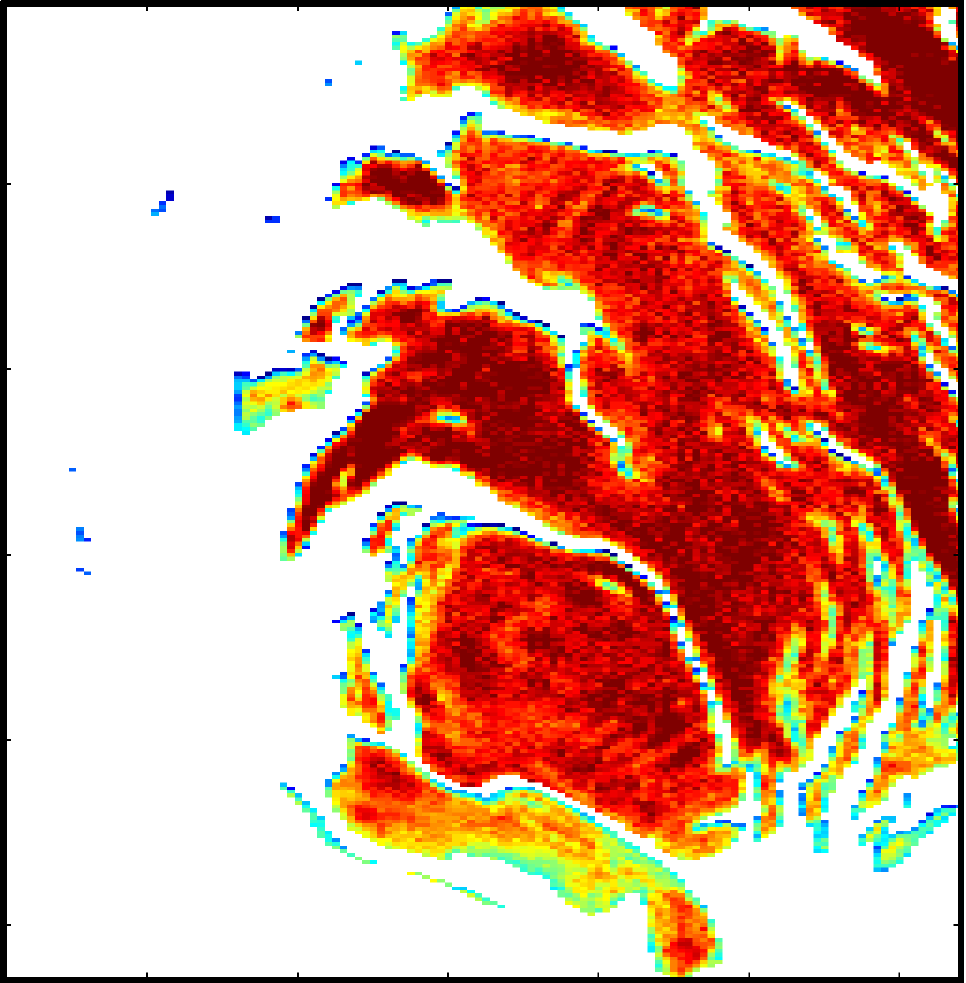}
    \includegraphics[width=.95\linewidth,clip=,keepaspectratio]{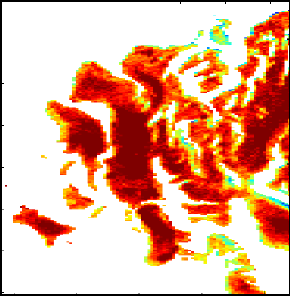}
    \\ RVM
   \end{minipage} \hfill
   \begin{minipage}[c]{.200\linewidth}
    \centering
    \includegraphics[width=.95\linewidth,clip=,keepaspectratio]{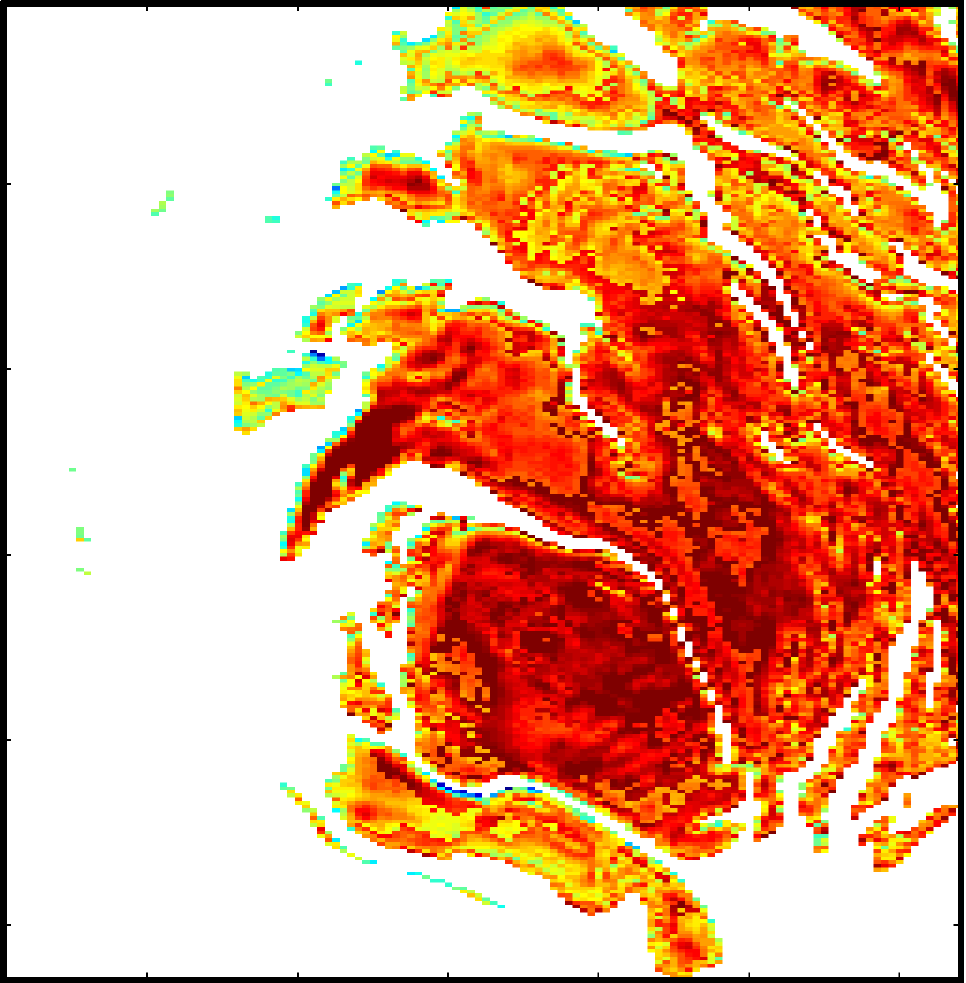}
    \includegraphics[width=.95\linewidth,clip=,keepaspectratio]{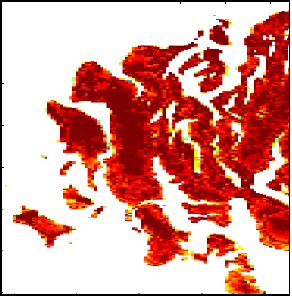}
     \\ MLE
   \end{minipage} \hfill
   \begin{minipage}[c]{.200\linewidth}
    \centering
    \includegraphics[width=.95\linewidth,clip=,keepaspectratio]{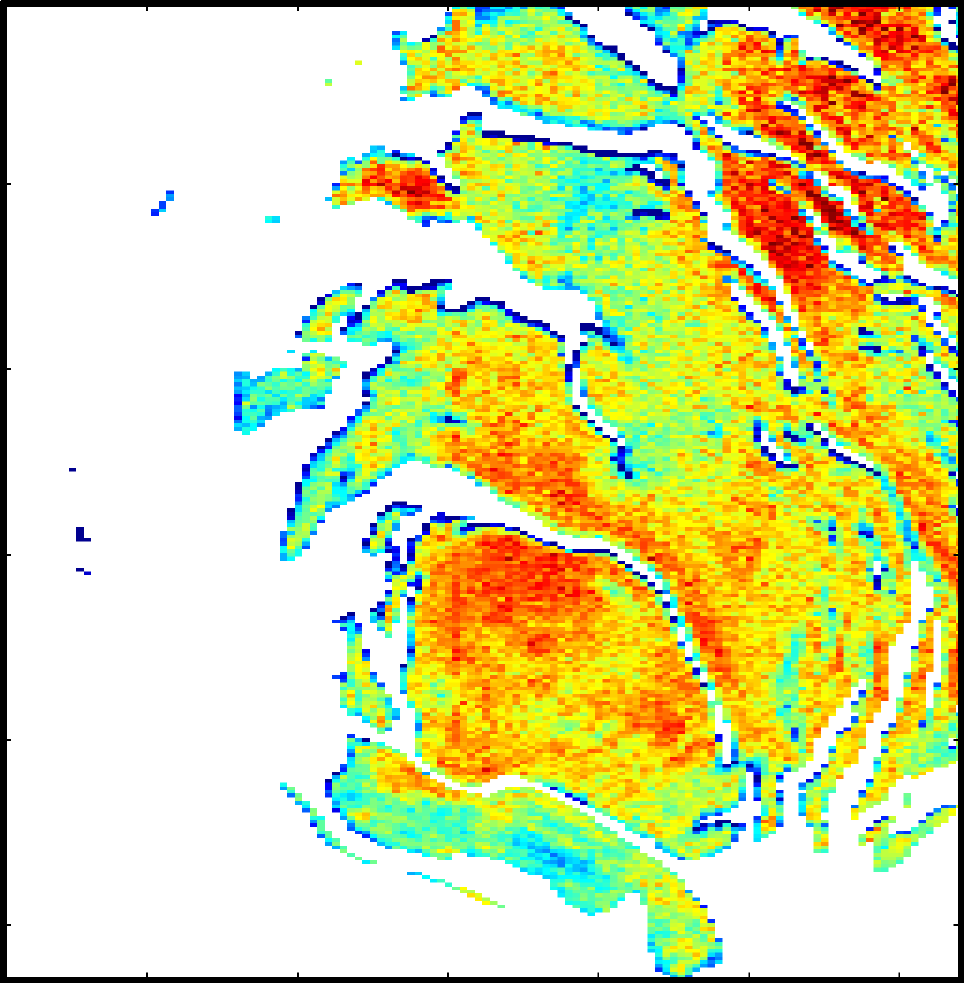}
    \includegraphics[width=.95\linewidth,clip=,keepaspectratio]{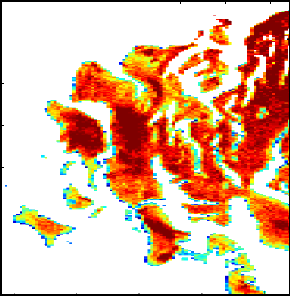}
     \\ JGMM
   \end{minipage} \hfill
   \begin{minipage}[c]{.074\linewidth}
    \centering
    \includegraphics[width=.95\linewidth,clip=,keepaspectratio]{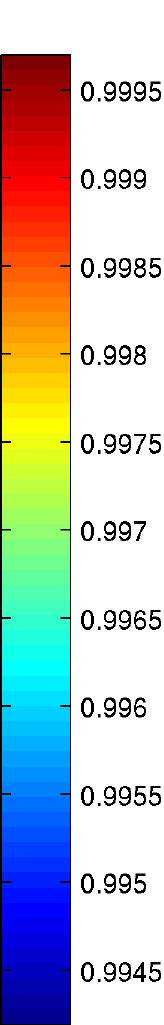}
    \\ $\mbox{ }$
   \end{minipage} \hfill    
   }
    \caption{\label{fig:mars} Proportions obtained with our method (hGLLiM-2) and three other methods. The data correspond to hyper-spectral images grabbed from two different viewpoints of the South polar cap of Mars. Top rows: orbit 41; Bottom rows: orbit 61. White areas correspond to unexamined regions, where the synthetic model does not apply.} 
\end{figure*}

\begin{figure*}
    \centering
    \begin{minipage}[c]{.190\linewidth}
    \centering
    \includegraphics[width=.95\linewidth,height=.70\linewidth,clip=]{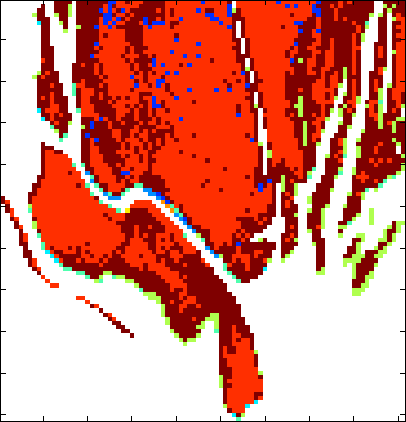}
    \includegraphics[width=.95\linewidth,height=.70\linewidth,clip=]{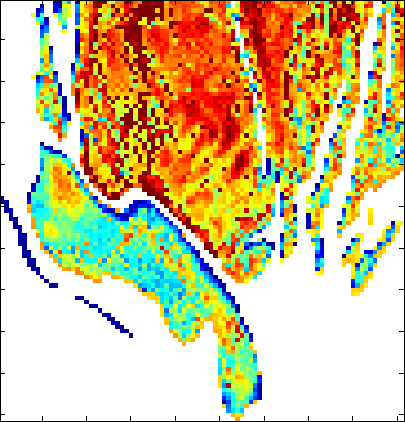}
    \includegraphics[width=.95\linewidth,height=.70\linewidth,clip=]{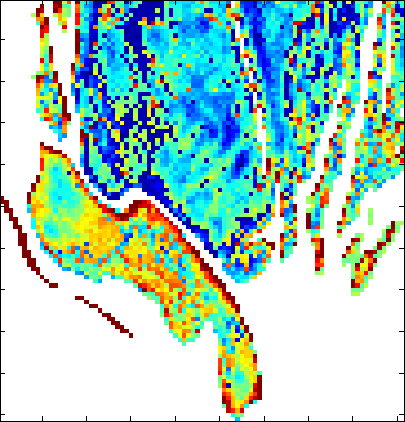}   
    \\ $\beta=0$
   \end{minipage} \hfill
   \begin{minipage}[c]{.190\linewidth}
    \centering
    \includegraphics[width=.95\linewidth,height=.70\linewidth,clip=]{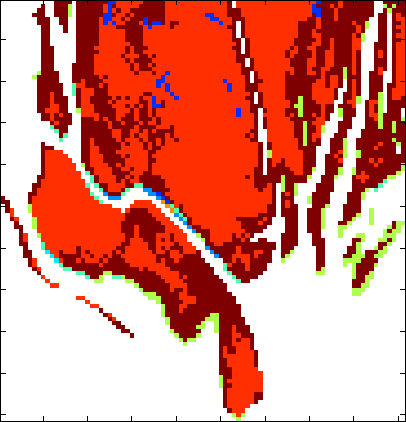}
    \includegraphics[width=.95\linewidth,height=.70\linewidth,clip=]{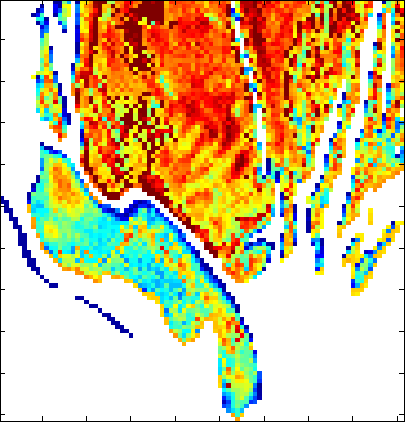}
    \includegraphics[width=.95\linewidth,height=.70\linewidth,clip=]{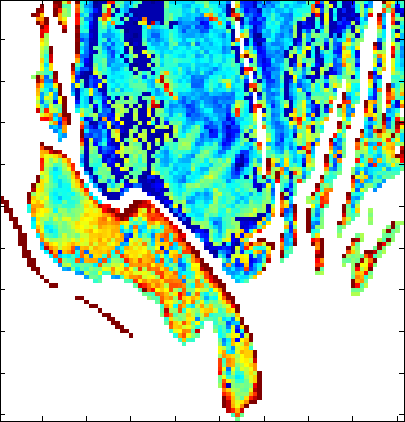}    
    \\ $\beta=10$
   \end{minipage} \hfill
   \begin{minipage}[c]{.190\linewidth}
    \centering
    \includegraphics[width=.95\linewidth,height=.70\linewidth,clip=]{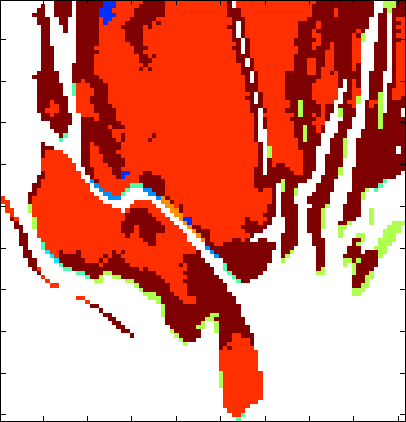}
    \includegraphics[width=.95\linewidth,height=.70\linewidth,clip=]{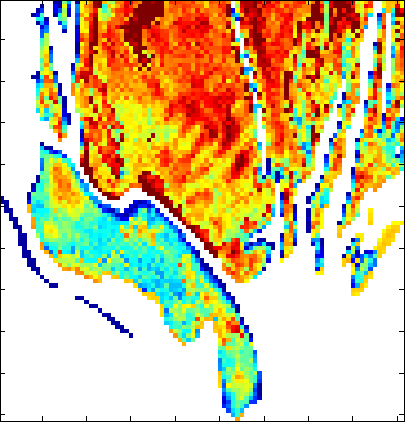}
    \includegraphics[width=.95\linewidth,height=.70\linewidth,clip=]{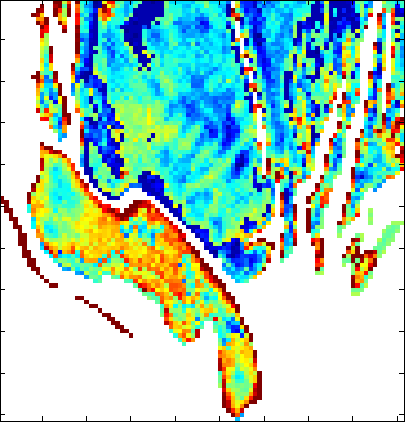} 
    \\ $\beta=50$
   \end{minipage} \hfill
   \begin{minipage}[c]{.190\linewidth}
    \centering
    \includegraphics[width=.95\linewidth,height=.70\linewidth,clip=]{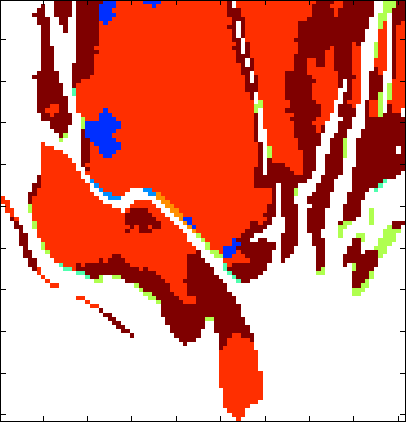}
    \includegraphics[width=.95\linewidth,height=.70\linewidth,clip=]{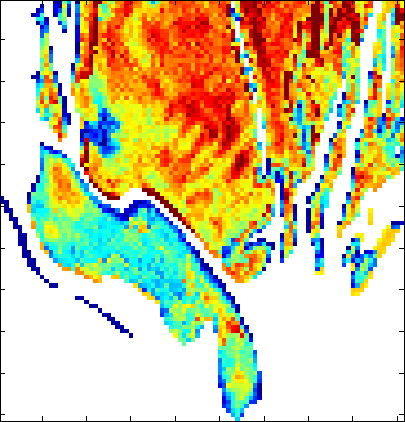}
    \includegraphics[width=.95\linewidth,height=.70\linewidth,clip=]{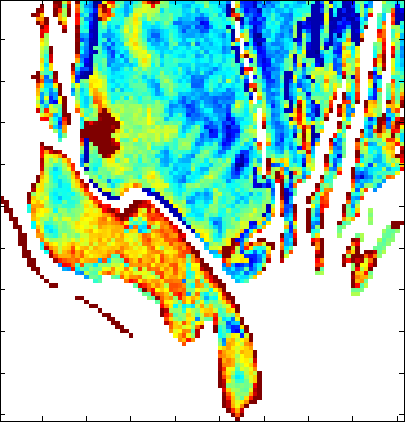}    
    \\ $\beta=100$
   \end{minipage} \hfill
   \begin{minipage}[c]{.190\linewidth}
    \centering
    \includegraphics[width=.95\linewidth,height=.70\linewidth,clip=]{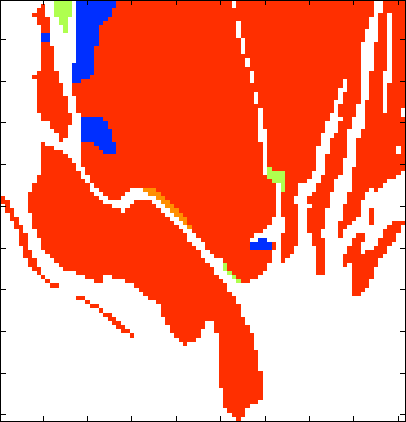}
    \includegraphics[width=.95\linewidth,height=.70\linewidth,clip=]{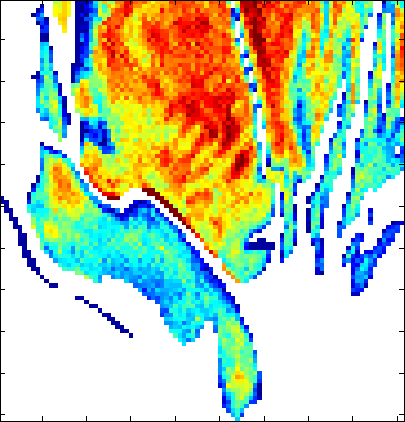}
    \includegraphics[width=.95\linewidth,height=.70\linewidth,clip=]{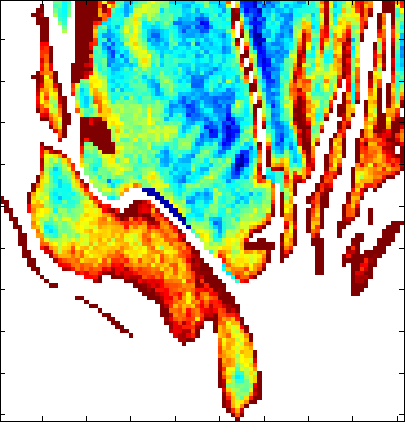}    
    \\ $\beta=500$
   \end{minipage} \hfill
     \caption{\label{fig:mars_beta_41} Qualitative results obtained with MRF-hGLLiM-2 for orbit~41. Top row: Each pixel $n$ is colored according to the class $k$ maximizing $p(Z_n=k|\Yvect_n=\yvect_n;\phivect)$ (each class index is assigned an arbitrary color). Middle row: Proportion of CO$_2$ ice. Bottom row: Proportion of Dust. Concentration color codes are the same as in Fig.~\ref{fig:mars}. To better appreciate the smoothing effect, a zoomed region is shown.} 
\end{figure*}
\begin{figure*}
   \begin{minipage}[c]{.190\linewidth}
    \centering
    \includegraphics[width=.95\linewidth,height=.70\linewidth,clip=]{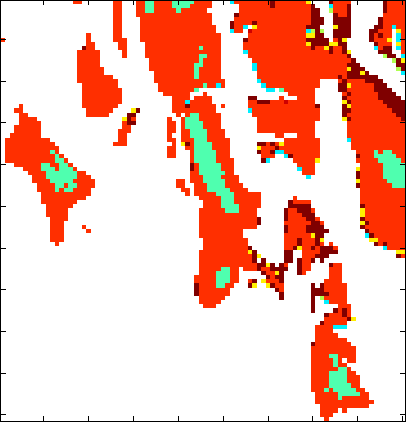}
    \includegraphics[width=.95\linewidth,height=.70\linewidth,clip=]{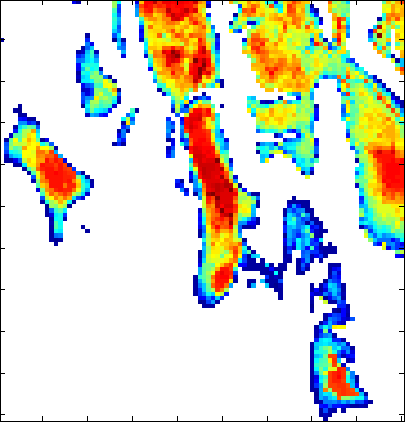}
    \includegraphics[width=.95\linewidth,height=.70\linewidth,clip=]{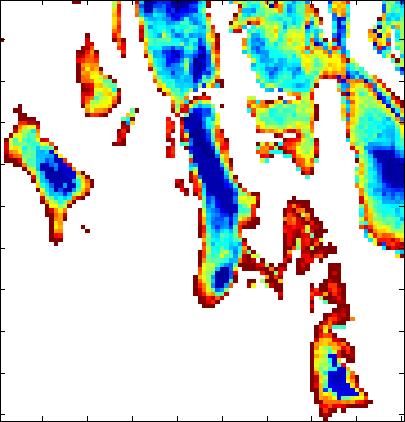}   
    \\ $\beta=0$
   \end{minipage} \hfill
   \begin{minipage}[c]{.190\linewidth}
    \centering
    \includegraphics[width=.95\linewidth,height=.70\linewidth,clip=]{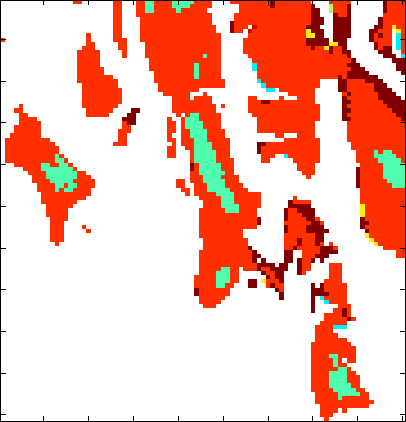}
    \includegraphics[width=.95\linewidth,height=.70\linewidth,clip=]{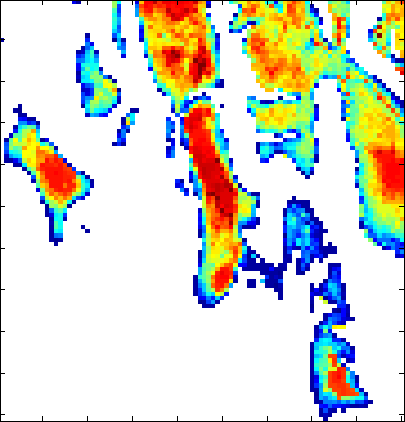}
    \includegraphics[width=.95\linewidth,height=.70\linewidth,clip=]{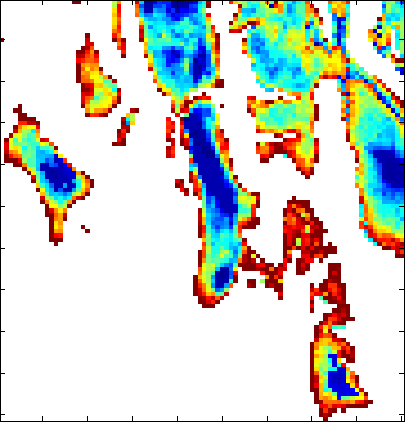}    
    \\ $\beta=10$
   \end{minipage} \hfill
   \begin{minipage}[c]{.190\linewidth}
    \centering
    \includegraphics[width=.95\linewidth,height=.70\linewidth,clip=]{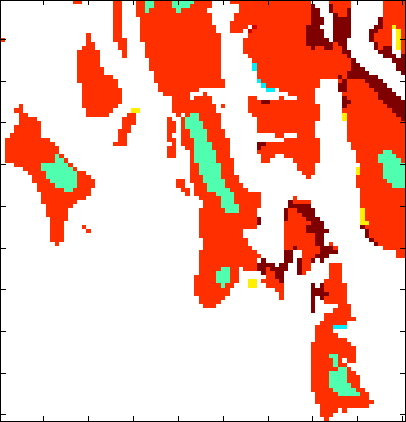}
    \includegraphics[width=.95\linewidth,height=.70\linewidth,clip=]{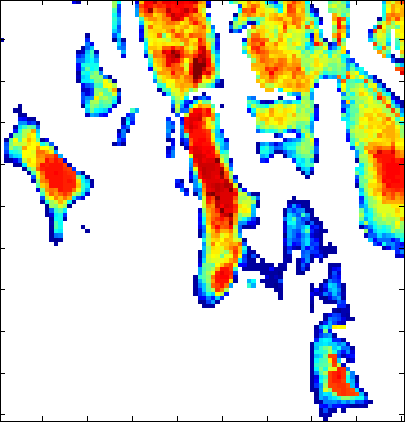}
    \includegraphics[width=.95\linewidth,height=.70\linewidth,clip=]{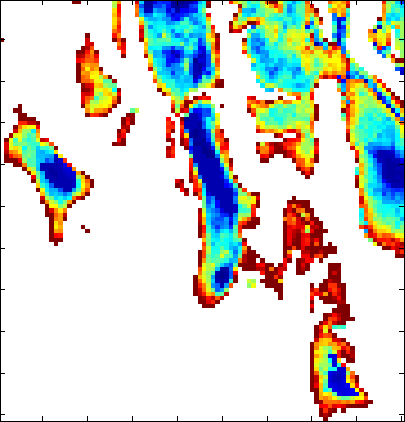} 
    \\ $\beta=50$
   \end{minipage} \hfill
   \begin{minipage}[c]{.190\linewidth}
    \centering
    \includegraphics[width=.95\linewidth,height=.70\linewidth,clip=]{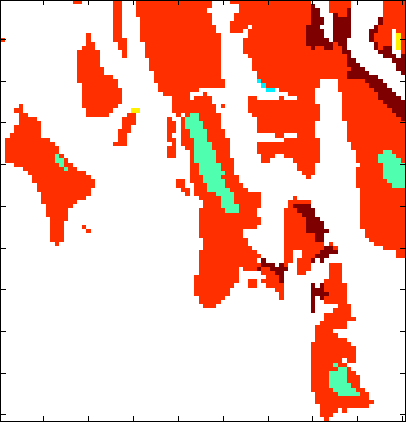}
    \includegraphics[width=.95\linewidth,height=.70\linewidth,clip=]{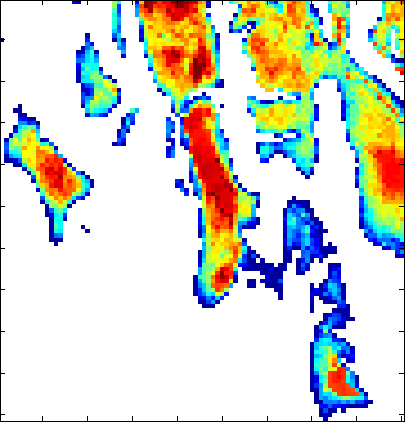}
    \includegraphics[width=.95\linewidth,height=.70\linewidth,clip=]{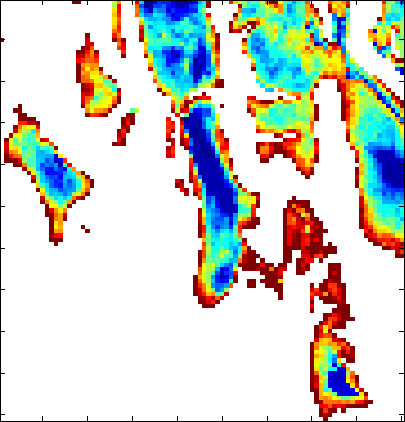}    
    \\ $\beta=100$
   \end{minipage} \hfill
   \begin{minipage}[c]{.190\linewidth}
    \centering
    \includegraphics[width=.95\linewidth,height=.70\linewidth,clip=]{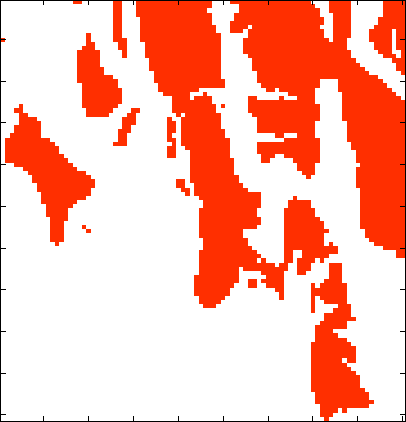}
    \includegraphics[width=.95\linewidth,height=.70\linewidth,clip=]{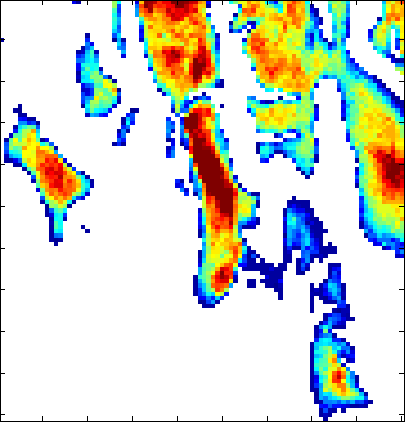}
    \includegraphics[width=.95\linewidth,height=.70\linewidth,clip=]{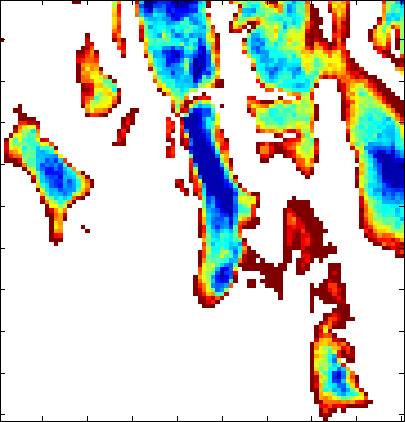}    
    \\ $\beta=500$
   \end{minipage} \hfill 
    \caption{\label{fig:mars_beta_61} Qualitative results obtained with MRF-hGLLiM-2 for orbit~61. Top row: Each pixel $n$ is colored according to the class $k$ maximizing $p(Z_n=k|\Yvect_n=\yvect_n;\phivect)$ (each class index is assigned an arbitrary color). Middle row: Proportion of CO$_2$ ice. Bottom row: Proportion of Dust. Concentration color codes are the same as in Fig.~\ref{fig:mars}. To better appreciate the smoothing effect, a zoomed region is showed for each orbit.} 
\end{figure*}
Table \ref{tab:planeto_err} shows NRMSE values obtained for the three remaining parameters. The ground-truth latent variable dimension is $L^{*}_{\textrm{w}}=2$, and accordingly, the empirically best latent dimension for hGLLiM is $L^{\dagger}_{\textrm{w}}=2$. hGLLiM-2 outperforms all the other methods on that task, more precisely  the error is $36\%$ lower than the second best-performing method, RVM, closely followed by MLE. No significant difference is observed between hGLLiM-2 and hGLLiM-3. \addnote[complexity1b]{1}{The average training time of MLE was 12.2 seconds and the average training time of hGLLiM was 18.8 seconds using our Matlab implementation. Due to the computation of the $D\times D$ kernel matrix, the training time of RVM was about 10 times larger.}

\addnote[bic]{1}{We also tested the selection of the latent dimension ($L_{\textrm{w}}$) based on BIC (Section \ref{subsec:BIC}).  For each training set, the hGLLiM-BIC method minimizes BIC with the latent dimension in the range $0\le L_{\textrm{w}}\le20$, and used the corresponding model to perform the regression. Interestingly, hGLLiM-BIC performed very well on these large training sets ($N=10,000$) as it correctly selects $L_{\textrm{w}}=2$ for the 20 training sets used by BIC (\textcolor{black}{the BIC selection could differ with the training set}), yielding the same results as those obtained with hGLLiM-2}. Interestingly, notice how hGLLiM and almost all the methods benefit from removing the ``faulty'' parameters that are difficult to estimate. In all cases, a slight performance increase for the reconstruction of the remaining three others parameters is observed when comparing to the results of Table \ref{tab:planeto_full_err}.

\subsection{\addnote[synth_images]{1}{MRF-hGLLIM Evaluation on Synthetic Hyper-spectral Images}}
\label{sec:synth_images}
\label{subsection:synth_images}
 
 \addnote[numeval1]{1}{
The data used in the previous section is not spatial, in the sense that there is no dependency between the different vectors of observations. In order to assess the advantage of the proposed MRF extension of hGLLiM when spatial dependencies exist, we synthesized 50 hyper-spectral images from the synthetic spectra dataset presented in the previous section. We generated $400\times300$ pixel images consisting of 12 ($4\times3$) square regions (see Figure \ref{fig:synth_spatial}), where each region corresponds to a set of similar underlying physical parameters. More specifically, for each region, a vector of physical parameters in $\mathbb{R}^{L_t}$ was picked, and the region was filled with this vector as well as its 15 nearest neighbors in the dataset. A synthetic hyper-spectral image was then generated from associated spectra, and corrupted with additive white Gaussian noise with a signal-to-noise-ratio of 6 dB.}

\addnote[numeval2]{1}{
We first learned the model parameters $\hat{\thetavect}$ using hGLLiM-0 (equivalent to MLE \cite{XuJordanHinton95}) and 10,000 individual spectra associated to all 5 physical parameters, as explained in the previous section. We then compared the performance of MRF-hGLLiM-0 when $\beta$ was automatically estimated from the test images (section \ref{subsec:inverseMRF}) and when $\beta$ was set to 0, \textit{i.e.} no spatial dependency, which corresponds to hGLLiM-0. The spectra used in the 50 test synthetic images were outside the training set. The MRF prior was incorporated using a 8-pixel neighborhood and MRF-hGLLiM-0 estimated $\beta$ values comprised between 2.8 et 3.1 for these synthetic images. Table \ref{tab:planeto_spatial_err} shows the NRMSE obtained with both methods. The use of spatial dependencies significantly reduced the estimation errors for all concentrations, in all 50 test images.  For each image, the averaged across pixels  errors were computed for both hGLLiM-0 and MRF-hGLLiM-0 and the corresponding two samples of size 50 were compared using a paired t-test. The test confirmed a highly signficant gain in favor of MRF-hGLLiM-0.} \addnote[complexity2]{1}{The average computational time for a $100\times 100$ test image  is 220s using hGLLiM-0 and 550 using MRF-hGLLiM-0 with our Matlab implementation}.

\subsection{Real Mars Hyper-spectral Images}
We now investigate the potential of the proposed
MRF-hGLLiM method using a dataset of real hyper-spectral images
collected from the imaging spectrometer OMEGA instrument
\cite{Bibring2004} on board of the Mars Express spacecraft. We adopt the evaluation protocol proposed in 
 \cite{bernard2009retrieval}. An adequately
selected subset of 6,983 spectra of the synthetic database is used for model training. 
Then, trained models are applied to the real data made of observed spectra acquired from satellite hyper-spectral sensors.
For evaluation, we focus on data from Mars's South polar cap.
Since  ground-truth for the physical properties of Mars's South pole regions is not currently available,  we consider a qualitative
evaluation by comparing  five models: hGLLiM-2, MRF-hGLLiM-2, and the three best performing methods,
among the tested ones, namely RVM, MLE and JGMM.

First, hGLLiM-2, RVM, MLE and JGMM are used to extract physical parameters from 
two hyper-spectral images of the same area but seen from two slightly different view points (orbit
41 and orbit 61). Since responses are proportions  with values between 0
and 1, values smaller than 0 or higher than 1 are not admissible and hence they are set to either one of the closest bounds. In the real data setting, the best performance is achieved when no data mean-variance  normalization is performed, and using $K=10$ for hGLLiM-2, MLE and JGMM. 
As it can be seen in 
Fig.~\ref{fig:mars}, hGLLiM-2 estimates
proportion maps with similar characteristics for the two view
points, which suggests a good model-response consistency. Such a consistency is not
observed with the other three tested methods. Furthermore, RVM and
MLE result in a much larger number of values outside the
admissible interval $[0,1]$. Moreover, hGLLiM-2 is the only method featuring
less dust at the South pole cap center and higher proportions
of dust at the boundaries of the CO2 ice, which matches expected
results from planetology \cite{Doute2005}. Finally, note that the
proportions of CO2 ice and dust clearly seem to be complementary
using hGLLiM-2, while this complementarity is not obviously noticed when using
the other methods.

Despite these satisfying qualitative results, hGLLiM-2 produces less regular maps than the other methods. This is particularly true for orbit 41, as can be seen in the top rows of Fig.~\ref{fig:mars}(a) and (b). This does not match physical expectations since chemical proportions should vary rather smoothly in space. To address this issue, we use MRF-HGLLiM-2 by exploiting the MRF-forward mapping (\ref{eq:JGMM_inverse_exp_MRF}).

The MRF-hGLLiM-2 model is trained  using synthetic data without accounting for the spatial dependencies ($\beta$ set to $0$). This training results in a model equivalent to hGLLiM-2. However, during testing, the MRF prior is enforced using an 8-pixel neighborhood.  For illustration purpose, the parameter $\beta$ is empirically set to control spatial smoothness of the resulting physical parameter map. According to the MRF prior model, the larger the value of $\beta$, the smoother the resulting parameter field. 

We compare the outputs of hGLLiM-2 and MRF-hGLLiM-2 with different interaction parameter values $\beta$. The top rows of Fig.~\ref{fig:mars_beta_41} and Fig.~\ref{fig:mars_beta_61} color each pixel according to the class $k$ maximizing $p(Z_n=k|\Yvect_n=\yvect_n;\phivect)$. By progressively increasing $\beta$ from 0 to 500, isolated points are removed, thus creating larger and larger image regions associated to a single class. As it can be seen in Fig.~\ref{fig:mars_beta_41}, this has the desired smoothing effect on the resulting proportion maps. Interestingly, MRF-hGLLiM-2 yields smoother maps than hGLLiM-2 while preserving its desirable qualitative performances, \textit{e.g.}, higher proportions of dust at boundaries and complementarity of CO2 and dust proportions. Notice that $\beta$ has a less significant effect on proportions obtained at orbit 61, \textit{e.g.}, Fig.~\ref{fig:mars_beta_61}, since hGLLiM-2 already yields quite smooth maps for this orbit. 

\section{Conclusions}
\label{sec:conclusions}
\vspace{-2mm}
In this paper, we proposed a Gaussian mixture of locally-affine regression model that maps a high-dimensional space of hyper-spectral observations onto a low-dimensional  space of physical parameters. Spatial regularity is enforced and integrated into the model through an MRF prior on the Gaussian-mixture hidden variables.
Model learning is achieved via a variational expectation-maximization procedure that is fully derived and described in detail. Model evaluation is conducted using both simulated data and real hyper-spectral images from the Mars Express satellite sensors. The proposed approach outperforms four state-of-the-art regression methods. Furthermore, when applied to actual images of the Mars South pole, our model produces spatially-regular and smooth maps in accordance with the MRF constraints. 

As a first model extension, we will investigate its application to the hyper-spectral un-mixing problem. In this context, we will design experimental setups to study the ability of the model to partially estimate hyper-spectral mixture components together with their corresponding proportions.
Further research, in line with dynamical geophysical data-assimilation research \cite{SixianJST2013,DongJST2013}, will be to investigate generalizations of the model in order to process temporal sequences of observations. To this end, relevant dynamical models of physical phenomena driving hyper-spectral observation evolution will be investigated and integrated. Equipped with an MRF spatial prior and a temporal dynamical model, our formulation could be able to address missing-data interpolation problems which are ubiquitous in remote sensing \cite{SalbergTGRS2011}.

\bibliographystyle{IEEEtran}

\begin{IEEEbiography}[{\includegraphics[width=1in,
height=1.25in,clip,keepaspectratio]{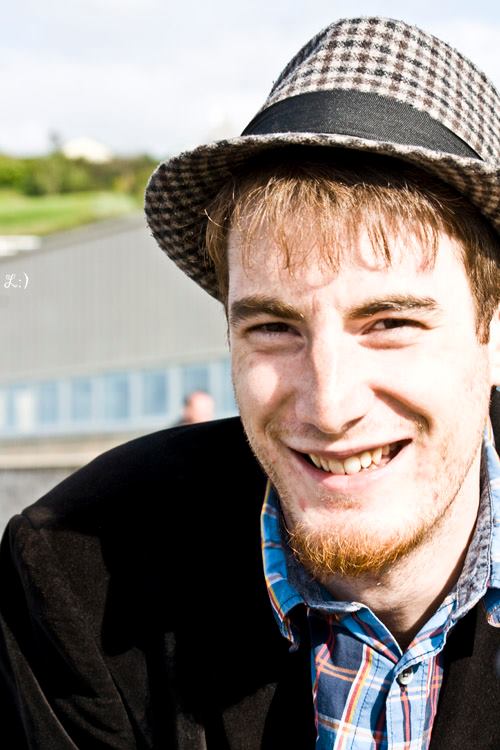}}]{Antoine Deleforge}
received the B.Sc. (2008) and M.Sc. (2010) engineering degrees in computer science and mathematics from the Ecole Nationale Sup\'erieure dÕInformatique et Math\'ematiques Appliqu\'ees de Grenoble (ENSIMAG), France, as well as the specialized M.Sc. (2010) research degree in computer graphics, vision, robotics from the Universit\'e Joseph Fourier, Grenoble. In 2013, he received the Ph.D. degree in computer science and applied mathematics from the University of Grenoble. Since January 2014, he has as a post-doctoral fellow appointment with the chair of Multimedia Communication and Signal Processing of the Erlangen-Nuremberg University, Germany. His research interests include machine learning for signal processing, Bayesian statistics, computational auditory scene analysis, and robot audition.
\end{IEEEbiography}

\begin{IEEEbiography}[{\includegraphics[width=1in,
height=1.25in,clip,keepaspectratio]{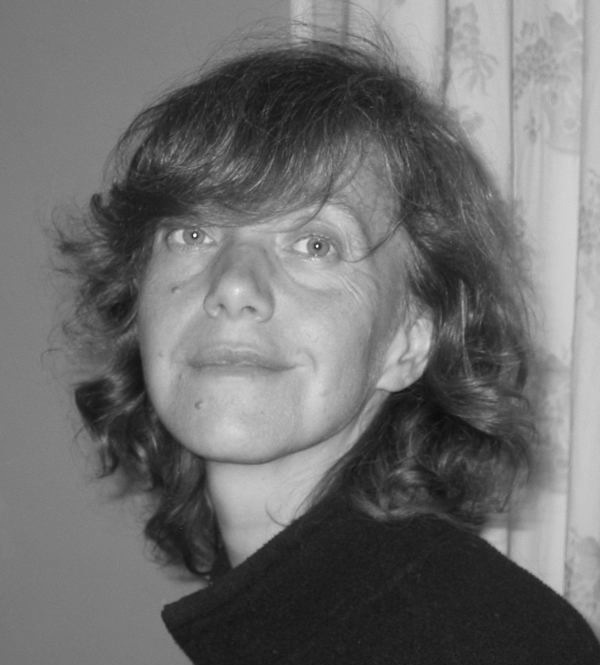}}]{Florence Forbes}
received the B.Sc. and M.Sc.
degrees in computer science and applied
mathematics from the Ecole Nationale Sup\'erieure
dÕInformatique et Math\'ematiques Appliqu\'ees de Grenoble (ENSIMAG), France, and the PhD
degree in applied probabilities from the University
Joseph Fourier, Grenoble, France. Since
1998, she has been a research scientist with the
Institut National de Recherche en Informatique
et Automatique (INRIA), Grenoble Rh\^one-Alpes,
Montbonnot, France, where she founded the MISTIS team and has been the team head
since 2003. Her research activities include Bayesian analysis,
Markov and graphical models, and hidden structure models.
\end{IEEEbiography}

\begin{IEEEbiography}[{\includegraphics[width=1in,
height=1.25in,clip,keepaspectratio]{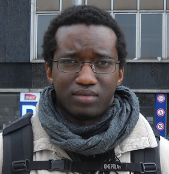}}]{Sil\`{e}ye Ba}
received the M.Sc. (2000) in applied mathematics and signal processing from University of Dakar, Senegal, and the M.Sc. (2002) in mathematics, computer vision, and machine learning from Ecole Normale Sup\'erieure de Cachan, Paris, France. From  
2003 to 2009 he was a PhD student and then a post-doctoral researcher at IDIAP Research Institute, Martigny, Switzerland
where he worked on probabilistic models for object tracking and human activity recognition. From 2009 to 2013, he was a researcher at Telecom Bretagne, Brest, France working on variational models for multi-modal geophysical data processing. From 2013 to 2014 he worked at RN3D Innovation Lab, Marseille, France, as a research engineer, where he used computer vision and machine learning principles and methods to develop human-computer interaction software tools. Since June 2014 he has been a researcher in the Perception
team at INRIA Grenoble Rh\^one-Alpes working on machine learning and computer vision models for natural human-robot interaction. 
\end{IEEEbiography}

\begin{IEEEbiography}[{\includegraphics[width=1in,
height=1.25in,clip,keepaspectratio]{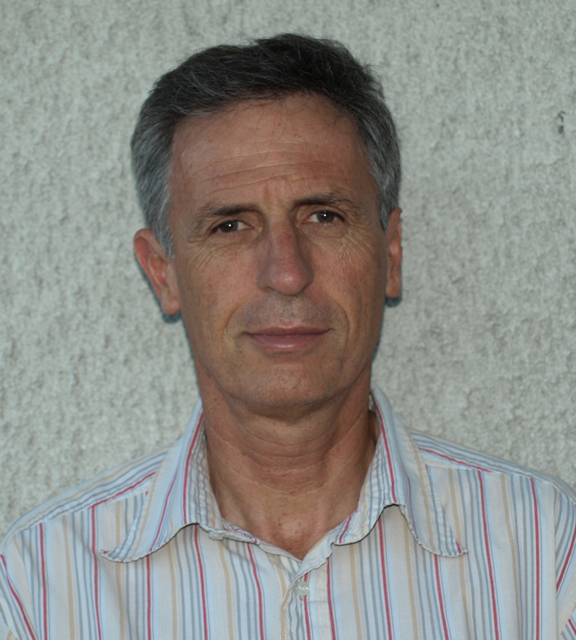}}]{Radu Horaud} 
received the B.Sc. (1977) degree in electrical engineering, the M.Sc. (1979) degree
in control engineering, and the Ph.D. (1981) degree in computer science from
the Institut National Polytechnique de Grenoble, France. 
Currently he holds a position of director of research with the Institut National de Recherche en Informatique et
Automatique (INRIA), Grenoble Rh\^one-Alpes, Montbonnot, France, where
he is the founder and head of the PERCEPTION team. His
research interests include computer vision, machine learning, audio signal processing, 
audiovisual analysis, and robotics and human-robot interaction. He is an area editor of the
\textit{Computer Vision and Image Understanding}, a member of
the advisory board of the \textit{International Journal of Robotics
  Research}, and an associate editor of the
\textit{International Journal of Computer Vision}. In 2013, Radu Horaud was awarded a five year ERC Advanced Grant for his project \textit{Vision and Hearing in Action} (VHIA, 2014-2019).
\end{IEEEbiography}

\end{document}